\documentclass[preprint,showpacs,preprintnumbers,amsmath,amssymb,superscriptaddress]{revtex4}
\usepackage{graphicx,color}
\usepackage{amsmath,amssymb}
\usepackage{url}
\usepackage{epstopdf}
\newcommand{\hs}{\hspace*{0.5cm}}

\newcommand{\be}{\begin{equation}}
\newcommand{\ee}{\end{equation}}
\newcommand{\bea}{\begin{eqnarray}}
\newcommand{\eea}{\end{eqnarray}}
\newcommand{\nn}{\nonumber}
\newcommand{\crn}{\nonumber \\}

\newcommand{\fr}{\frac}

\newcommand{\bc}{\begin{center}}
\newcommand{\ec}{\end{center}}

\newcommand {\ba}{\begin{array}}
\newcommand {\ea}{\end{array}}
\newcommand{\ben}{\begin{enumerate}}
\newcommand{\een}{\end{enumerate}}

\usepackage{bm}
\usepackage{dcolumn}

\begin{document}

\title{ Lepton flavor violating decay of SM-like Higgs  in a radiative  neutrino mass  model }
\author{ T. T. Thuc}\email{ttthuc@grad.iop.vast.vn}
\affiliation{Institute of Physics,   Vietnam Academy of Science and Technology, 10 Dao Tan, Ba
Dinh, Hanoi, Vietnam }
\author{L. T. Hue}\email{lthue@iop.vast.vn}
\affiliation{Institute of Physics,   Vietnam Academy of Science and Technology, 10 Dao Tan, Ba
Dinh, Hanoi, Vietnam }
\author{H. N. Long}\email{hoangngoclong@tdt.edu.vn}
\address{Theoretical Particle Physics and Cosmology Research Group, Ton Duc Thang University, Ho Chi Minh City, Vietnam}
\address{Faculty of Applied Sciences,
 Ton Duc Thang University, Ho Chi Minh City, Vietnam}
\author{T. Phong Nguyen}\email{thanhphong@ctu.edu.vn}
\affiliation{Department of Physics, Cantho University, 3/2 Street, Ninh Kieu, Cantho, Vietnam}

\begin{abstract}
The lepton flavor violating decay of the Standard Model-like Higgs (LFVHD) is discussed in the framework of the radiative neutrino mass model built in \cite{Kenji}. The branching ratio (BR) of the LFVHD are shown to reach $10^{-5}$ in the most interesting region of the parameter space shown in \cite{Kenji}. The dominant contributions come from the singly charged Higgs mediations, namely the coupling of $h^\pm_2$ with exotic neutrinos. Furthermore, if  doubly charged Higgs is heavy enough to allow the  mass of $h^\pm_2$ around 1 TeV, the mentioned BR can reach $10^{-4}$. Besides, we have obtained that the large values of the Br$(h\rightarrow\mu\tau)$ leads to very small ones of the Br$(h\rightarrow e\tau)$,  much smaller than various sensitivity of current experiments.
\end{abstract}
\pacs{
12.60.Fr, 13.15.+g,  14.60.St, 14.80.Bn
 }
\maketitle
\section{Introduction}
\label{sec:intro}
The confirmation of the existence of  a scalar boson, known as the Standard  Model  (SM)- like Higgs, is  the greatest early success  of  the LHC \cite{higgsdicovery1,higgsdicovery2}. In addition, the LHC has  reported recently some significant new physics beyond the SM where the LFVHD $h\rightarrow\mu\tau$  is one of the hottest subjects \cite{exLFVh}.   The upper bound  Br$(h^0\rightarrow\mu\tau) < 1.5\times 10^{-2}$ at 95\% C.L. was announced by the  CMS Collaboration, in agreement with $1.85\times 10^{-2}$ at 95\% C.L. from the ATLAS Collaboration.  More interestingly, the CMS has indicated a $2\sigma$ excess of this decay, with the value of Br$(h\rightarrow\mu\tau)=0.84^{+0.39}_{-0.37}\%$. Besides, two other lepton flavor violating (LFV) decays of the SM-like Higgs have set experimental upper bounds  at  BR$(h\rightarrow e\tau)<0.7\%$ and BR$(h\rightarrow e\mu)<0.036\%$ at $95\%$ C.L.\cite{LFVhtauemu}.  Theoretically, many publications have studied how large the BR$(h\rightarrow\mu\tau)$ can become in specific models beyond the SM,  such as the seesaw \cite{seesaw,iseesaw}, supersymmetry (SUSY)\cite{seesaw,SUSY}, two Higgs Doublet \cite{THDL1,THDL2}, and 3-3-1 models \cite{LFVHDUgauge}, as well as other interesting ones \cite{NonSUSY,leptoquark,LFVgeneral}. The LFV decay of new neutral Higgs bosons in non-SUSY models has also been discussed \cite{lfvNH}. The significance  of LFVHD in colliders was addressed in \cite{LFVcol}.

The first source of LFV decays  results from the mixing of different flavor massive neutrinos \cite{NuOcs}.  The simplest models explaining the mixing and masses of active neutrinos may be the seesaw models, but the BR of LFV decays predicted by these models are very small. Perhaps the inverse seesaw model gives the largest BR, which is about  $10^{-5}$ \cite{iseesaw}.  All of the SUSY models, even the Minimal Supersymmetric Standard Model, easily predict large values of the BR of LFVHD with new LFV sources in the slepton sector.  However, the particle spectra of these models are rather complicated. In contrast, recent studies have shown that many of non-SUSY models inheriting simpler particle  spectra can predict very large BRs of the LFVHD at one loop level, satisfying all relevant experimental constraints.  Some of these models even have tree level couplings of LFVHD, and they simultaneously explain other interesting experimental results \cite{THDL1}.

There is another class of  models, where neutrino mass is radiatively generated, that can predict large BRs of the LFVHD. These models do not have active neutrino mass terms at tree level, but they contain LFV couplings of new particles such as scalars and new leptons in order to generate neutrino masses from loop contributions. There is an interesting property that the loop suppression factors appearing in the expression of neutrino masses lead to the alleviation of the hierarchy in couplings. Hence the aforementioned models will allow large Yukawa couplings, which may result in large  BR  values for many LFV processes. By investigating a specific model with three loop neutrino mass introduced in \cite{Kenji}, we try to make clear how large the BR of LFVHD can reach in the allowed regions. Furthermore, the contributions from active neutrino mediations may be enhanced  because the Glashow-Iliopoulos-Maiani (GIM) mechanism does not work. Using the 't Hooft-Feynman gauge where loop contributions from private Feynman diagrams are all finite, we can compute and compare them. As a result, the best regions for large BRs of the  LFVHD can be found with precise conditions of free parameters. The contributions from active neutrino mediations are divided completely into independent  contributions of  $W^\pm$ and new singly charged Higgs bosons.  As we will see later, the active neutrino loops in the radiative neutrino mass model may give significant contributions to LFV processes. This is different from all of the other models, where these contributions are either ignored; or  are difficult to estimate  when active neutrinos mix with new leptons, as in the case of the (inverse) seesaw models.

Our paper is arranged as follows. Section \ref{Rev} will collect all needed ingredients for calculating the BR of the LFVHD. Section \ref{para} concentrates on detailed  expressions of the LFVHD amplitudes and partial widths.  The constraints given in \cite{Kenji} will be discussed to find the  allowed regions of parameter space. A mumerical discussion is conducted  and the main results are summarized in  Secs. \ref{numerical} and \ref{Con}. Finally, the Appendices \ref{PVfunction} and \ref{ffactor} list analytic expressions of Passarino-Veltman (PV) functions and LFVHD form factors.  The divergence cancellations of particular one-loop Feynman diagrams in the t' Hooft-Feynman gauge  are proved in Appendix  \ref{ffactor}.

\section{\label{Rev}Review  of the  model}
\subsection{Particle content}
Following  Ref. \cite{Kenji}, the particle content of the model is listed in Table \ref{tab:1}, where the last row represents charges  of  an additional global symmetry,  $U(1)$.   Aside from the SM particles, new particles are all  gauge singlets,  including three Majorana fermions, $N_{R_{1,2,3}}$; one neutral Higgs boson $\Sigma_0$; four singly charged Higgs bosons $(h^\pm_1,h^\pm_2)$; and two doubly charged Higgses, $k^{\pm\pm}$.  After the breaking of the $U(1)$ symmetry, a remnant $Z_2$ symmetry keeps $N_{R_{1,2,3}}$ and $h^\pm_2$ as negative parity particles. The remaining particles are trivial. An interesting consequence is that the lightest Majorana neutrino, which has negative parity, will be stable and can be a dark matter candidate.
\begin{widetext}
\begin{center}
\begin{table}[h]
\begin{tabular}{|c|c|c|c|c|c|c|c|c|}\hline
&\multicolumn{3}{c|}{Lepton Fields} & \multicolumn{5}{c|}{Scalar Fields} \\\hline
{Characters} & ~$L'_{L_i}= \left(
                            \begin{array}{c}
                              \nu'_{L_i} \\
                              e_{L_i}\\
                            \end{array}
                          \right)$~ & ~$e_{R_i}$~ &~$N_{R_i}$~ & ~$\Phi$~ &
 ~$\Sigma_0$~ & ~$h^+_1$~  & ~$h^{+}_2$~ & ~$k^{++}$~ \\\hline
$SU(3)_C$ & $\bm{1}$ & $\bm{1}$ & $\bm{1}$ & $\bm{1}$ & $\bm{1}$ & $\bm{1}$ & $\bm{1}$ & $\bm{1}$ \\ \hline
$SU(2)_L$ & $\bm{2}$ & $\bm{1}$&  $\bm{1}$&$\bm{2}$ & $\bm{1}$ &$\bm{1}$  &$\bm{1}$  &$\bm{1}$ \\\hline
$U(1)_Y$ & $-1/2$ & $-1$ & $0$ & $1/2$  & $0$  & $1$  & $1$ & $2$ \\\hline
{$U(1)$} & $0$ & $0$ & $-x$ & $0$  & $2x$   & $0$ & $x$ & $2x$  \\\hline
\end{tabular}
\caption{Lepton and scalar fields proposed in \cite{Kenji}. The notations $L'_{L_i}$ and  $\nu'_{L_i}$ denote the flavor states in distinguish  with the
mass state $\nu_{L_i}$ used later.}
\label{tab:1}
\end{table}
\end{center}
\end{widetext}

The Yukawa sector $\mathcal{L}_{Y}$ respecting all mentioned symmetries  is given as
\be {-} \mathcal{L}_{Y}
=
(y_\ell)_{ij} \overline{ L'_{L_i}} \Phi e_{R_j}  + \frac12 (y_{L})_{ij}  \overline{(L'_{L_i})^c} L'_{L_j} h^+_1  + (y_{R})_{ij}  \overline{N_{R_i}} (e_{R_j})^c h_2^{-}   +  \frac12 (y_N)_{ij} \Sigma_0  \overline{(N_{R_i})^c} N_{R_j} 
+\rm{h.c.} {.}\label{Yukawa} \ee
In addition, when symmetries are broken   an effective Yukawa term appears after we take into account the loop contributions for generating active neutrino masses, namely,
 \be  {-} \mathcal{L}^{\mathrm{eff}}_{Y}= \frac{(m_{\nu})_{ab}}{\sqrt{2}v'} \times \overline{(\nu_{L_a})^c}\nu_{L_b} \times \Sigma_0, \label{effYukawa}\ee
 corresponding to the  active neutrino mass term derived in \cite{Kenji}.

The Higgs potential is
\bea 
\mathcal{V}
&=&
 m_\Phi^2 |\Phi|^2 + m_{\Sigma}^2 |\Sigma_0|^2 + m_{h_1}^2 |h^+_1|^2  + m_{h_2}^2 |h_2^{+}|^2   + m_{k}^2 |k^{++}|^2
  \nn\\
&+ & \Bigl[
 \lambda_{11}  \Sigma_0^* h^-_1 h^-_1 k^{++} +  \mu_{22} h^+_2 h^+_2 k^{--}   + {\rm h.c.}
 \Bigr]
  +\lambda_\Phi |\Phi|^{4}
  + \lambda_{\Phi\Sigma} |\Phi|^2|\Sigma_0|^2
 +\lambda_{\Phi h_1}  |\Phi|^2|h^+_1|^2  
 \nn\\
& +&\lambda_{\Phi h_2}  |\Phi|^2|h^+_2|^2 +\lambda_{\Phi k}  |\Phi|^2|k^{++}|^2
 + \lambda_{\Sigma} |\Sigma_0|^{4} +
 \lambda_{\Sigma h_1}  |\Sigma_0|^2|h^+_1|^2
  +\lambda_{\Sigma h_2}  |\Sigma_0|^2|h^+_2|^2
   \nn\\&+&\lambda_{\Sigma k}  |\Sigma_0|^2|k^{++}|^2
  + \lambda_{h_1} |h_1^{+}|^{4}
  {+} \lambda_{h_1 h_2}  |h_1^{+}|^2|h^+_2|^2 +\lambda_{h_1 k}  |h_1^{+}|^2|k^{++}|^2\nn\\
&+& \lambda_{h_2} |h_2^{+}|^{4} + \lambda_{h_2 k}  |h_2|^2|k^{++}|^2
+ \lambda_{k} |k^{++}|^{4}.
\label{HP}\eea
The  scalar fields  are parameterized  as
\bea \Phi=\left(
           \begin{array}{c}
             G^+_{w} \\
              \frac{v+\phi+i G_z}{\sqrt{2}} \\
           \end{array}
         \right), \hs \Sigma_0=\frac{v'+\sigma}{\sqrt{2}}e^{iG/v'},
 \label{Nscalar} \eea
 where $v\simeq 246$ GeV, $v'$ is a new vacuum expectation value (VEV), and $G^{\pm}_{w}$ and $G_Z$ are Goldstone bosons of $W^{\pm}$ and $Z$ bosons, respectively.

\begin{figure}[h]
  \centering
\includegraphics*[width=15cm]{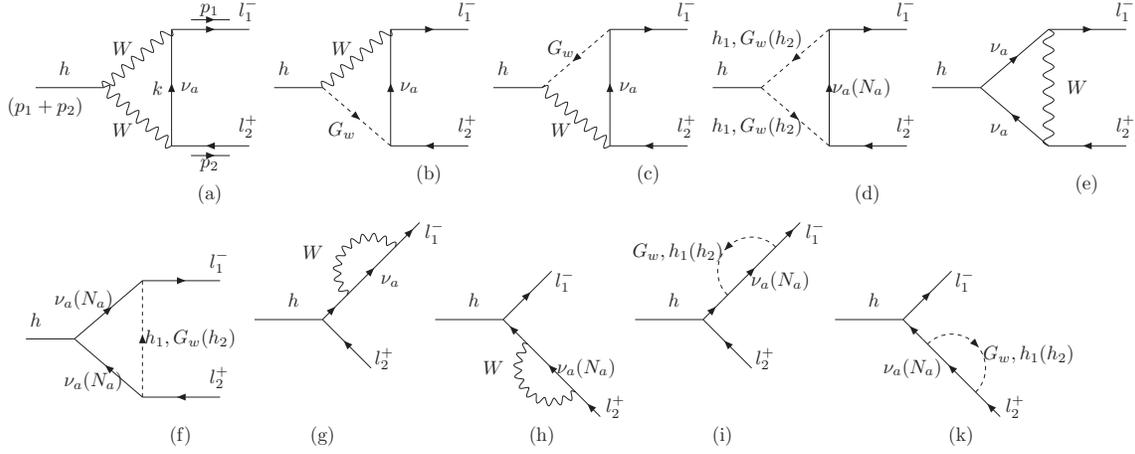}
 \caption{ Feynman diagrams for LEVHD $h\rightarrow \mu^{\pm}\tau^{\mp}$ decay, where $h$, $l_1\equiv e_2=\mu, l_2\equiv e_3=\tau$.  The parentheses imply that $N_a$ and  $h^\pm_2$  couple with each other only. }\label{LFVdiagram}
\end{figure}

 The above lepton and Higgs sectors show that the singly charged Higgs bosons  contribute mainly to the BR of the LFVHD while the doubly charged Higgs bosons do not. Because all of charged Higgs bosons are $SU(2)_L$ singlets, they do not couple with $W$ bosons like new Higgs  multiplets in other models. At the one loop level and t' Hooft-Feynman  gauge, the  Feynman diagrams for LFVHD are shown in Fig. \ref{LFVdiagram}.

 In order to investigate the LFVHD, the amplitudes will be calculated using the t' Hooft-Feynman gauge.  The amplitudes are formulated as functions of the PV functions analyzed  in \cite{LFVHDUgauge}. The notations of  four-component  (Dirac) spinors are used  for leptons. Specificially, the charged leptons are $e_i$, so that the left-handed component is $e_{L_i}=P_Le_i$   and right-handed one is $e_{R_i}=P_Re_i$, where $P_{R,L}\equiv(1\pm\gamma_5)/2$ are chiral operators. The corresponding charge  conjugations are $e^c_i{\equiv} C\overline{e_i}^T=\left(e^c_{L_i}, e^c_{R_i}\right)^T$, which satisfies $e^c_{L_i}=P_Le^c_i=C\overline{e_{R_i}}^T\equiv (e_{R_i})^c$ and $e^c_{R_i}=P_Re^c_i=C\overline{e_{L_i}}^T\equiv (e_{L_i})^c$. For the Majorana leptons such as active and exotic neutrinos, the  four-spinors are $\nu'_i=\nu'^c_i=\left( \nu'_{L_i},\; (\nu'_{L_i})^c\right)^T= \left( \nu'_{L_i},\; \nu^{\prime c}_{R_i}\right)^T$ and $N_i=N^c_i=\left( (N_{R_i})^c,\; N_{R_i}\right)^T =\left( N_{L_i}^c,\; N_{R_i}\right)^T$.  To reduce the second and third terms of (\ref{Yukawa}) to  more convenient forms, we will use  equalities like $\overline{e^c_{i}}P_L\nu'_{j}=  \overline{\nu'^c_{j}}P_Le_{i}=\overline{\nu'_{j}}P_Le_{i}$ and $\overline{N_{R_i}}(e_{R_j})^c=\overline{N_{i}}P_L e_j^c=\overline{ e_j}P_LN^c_{i}=\overline{ e_j}P_LN_{i}$   (Appendix G of \cite{spinor}).

\subsection{Mass spectrum and LFVHD couplings}
In the mass basis, the model consists of two CP-even neutral Higgs bosons; a  Nambu-Goldstone scalar boson $G$; four singly charged Higgs bosons, $h^{\pm}_1$, $h^{\pm}_2$; and two doubly charged Higgs bosons, $k^{\pm\pm}$. At one loop level, the LFVHD involves only two CP-even neutral and four singly charged Higgs bosons.  The two CP-even neutral Higgs bosons  are  a SM-like Higgs, $h$ ($m_h=125$ GeV) and a new CP-even one, $H$. They relate to the original Higgs components $(\phi,\sigma)$ via the transformation
\be  \left(
       \begin{array}{c}
         \phi \\
         \sigma \\
       \end{array}
     \right)=\left(
               \begin{array}{cc}
                 c_{\alpha} & s_{\alpha} \\
                 -s_{\alpha} & c_{\alpha} \\
               \end{array}
             \right)
      \left(
                \begin{array}{c}
                  h \\
                  H \\
                \end{array}
              \right),
 \label{Cnhiggs}\ee
 where $c_{\alpha}\equiv\cos \alpha,\; s_{\alpha}\equiv\sin \alpha,$ and $\alpha$ is defined as
 \be \sin2\alpha= \frac{2 \lambda_{\Phi\Sigma }vv'}{m^2_H-m^2_h}.\label{alp}\ee
 The masses $m_h$ and $m_H$ are  functions of the four parameters $\lambda_{\Phi},\lambda_{\Sigma},\lambda_{\Phi\Sigma}$ and $v'$. In the following calculations,
  we will fix $m_h=125.1$ GeV while  $m_H$, $ s_{\alpha}$ and $v'$ are taken as free parameters.  The original parameters are given as \cite{Kenji}
 \be \lambda_{\Phi\Sigma}=\frac{s_{\alpha}c_{\alpha}(m^2_H-m^2_h)}{vv'},\; \lambda_{\Phi}=\frac{c^2_{\alpha}m^2_h+ s_{\alpha}m^2_H)}{2v^2},\;  \mathrm{and}\; \lambda_{\Sigma}=\frac{s^2_{\alpha}m^2_h+ c^2_{\alpha}m^2_H}{2v'^2}.\label{laPS}\ee
 The perturbativity limit forces $\lambda_{\Phi},\lambda_{\Sigma},\lambda_{\Phi\Sigma}\leq 4\pi$, therefore gives an upper bound on $m_H$  with a large $s_{\alpha}$. For example,   $|s_{\alpha}|\leq 0.3$  corresponds to $m_H\leq 4$ TeV.

 The masses of the singly charged Higgs bosons are given as
 \be m^2_{h^{\pm}_1}=m^2_{h_1}+ \frac{1}{2}\left(\lambda_{\Phi h_1}v^2+ \lambda_{\Sigma  h_1}v'^2\right),\hs
 m^2_{h^{\pm}_2}=m^2_{h_2}+ \frac{1}{2}\left(\lambda_{\Phi h_2}v^2+ \lambda_{\Sigma  h_2}v'^2\right).\label{mschiggs}\ee
 Regarding the lepton sector, the first  and last terms in (\ref{Yukawa}) correspond to the mass terms of charged leptons and exotic neutrinos, respectively. They are assumed to be diagonal, i.e., the flavor basis and mass basis coincide. Their expressions are obtained as
 \bea  m_{e_i}=(y_\ell)_{ii}\frac{v}{\sqrt{2}}, \hs m_{N_i}= \frac{v'}{\sqrt{2}}(y_N)_{ii}. \label{mlepton1} \eea
 The  active neutrino masses generated from three loop corrections  can be expressed by the effective term (\ref{effYukawa}). The components of active neutrino mass matrix are given as \cite{Kenji}
 \begin{align}
(m_{\nu})_{ab}
&=
\frac{\mu_{11}\mu_{22}}{(4\pi)^6}
\sum_{i,j,k=1}^{3}
{\frac{1}{M_k^4}}
\left[
 (y_L)_{ai}m_{e_i}
 (y^T_{R})_{ik}
 (M_{N_k})
  (y_{R})_{kj}
 m_{e_j}
  (y^T_L)_{jb}
 \right] \notag \\
&\quad\quad
\times
F_1\left(
\frac{m_{h^+_1}^2}{{M_k^2}},
\frac{m_{h^+_2}^2}{{M_k^2}},
\frac{m_{\ell_i}^2}{{M_k^2}},
\frac{m_{\ell_j}^2}{{M_k^2}},
\frac{{M_{N_k}}^2}{{M_k^2}},
\frac{m_{k^{\pm\pm}}^2}{{M_k^2}}
\right),
	\label{eq:3loop_neutrino_mass}
\end{align}
where ${M_k}$ is the maximal value among the quantities $m_{h^\pm_1},m_{h^\pm_2},m_{e_{i}},m_{e_{j}},M_{N_k},m_{k^{\pm\pm}}$; $\mu_{11}\equiv\lambda_{11}v'/\sqrt{2}$; and $F_1$ is the three loop function given  in detail in \cite{Kenji}.

All of  the  mass  terms- together with couplings $h\overline{f}f$, where $f=e_i, \nu_i$ and $N_i$- are  parts of  Yukawa terms of neutral Higgs bosons. The flavor states $\nu'_{L_i}$ and mass  states $\nu_{L_i}$  ($i=1,2,3$) of active neutrinos are related by the transformation $\nu'_{L_i}=U^L_{ij}\nu_{L_j}$, where $U^LU^{L\dagger}=U^{L\dagger}U^L=1$. The masses and mixing angles of the active neutrinos
 are taken from the best-fit experimental data given in  \cite{actnuUpdate}. The only unknown parameter is the  lightest mass.

 Concerning only on the mass terms  and  couplings involving the LFVHD of the SM-like Higgs boson,  the Yukawa interactions (\ref{Yukawa})  and (\ref{effYukawa})  are written as follows:
\bea - \mathcal{L}'_{Y}&=& m_{e_i} \overline{e_i}e_i+\frac{1}{2} m_{\nu_i} \overline{\nu_i}\nu_i+ \frac{1}{2} m_{N_i} \overline{N_i}N_i+ \frac{m_{e_i}}{v}\overline{e_i}e_i \left( c_{\alpha}h  \right)
-
\left[\frac{m_{N_i}}{2v'}\overline{N_i}N_i + \frac{m_{\nu_i}}{2v'}\overline{\nu_i}\nu_i\right] s_{\alpha}h
\crn &+&
\frac{\sqrt{2}m_{e_i}}{v}\left[U^L_{ij}\overline{e_i}P_L\nu_jG^-_w + U^{L*}_{ij}\overline{\nu_j}P_Re_iG^+_w \right]
  +  (y_L^TU^{L*})_{ij}\overline{\nu_j}P_Le_i h^+_1 +   (y_L^TU^{L})_{ij}\overline{e_i}P_R\nu_j h^-_1\crn
 &+&
 (y_R^T)_{ij}\overline{e_i}P_LN_j h^-_2+  (y_R^T)_{ij}\overline{N_j}P_Re_i h^+_2.
 \label{Lyukawap}\eea
The LFV couplings relating  to the $W^{\pm}$ gauge boson only occur in the covariant kinetic terms of $SU(2)_L$ doublets, exactly the same as in the SM, namely,
\be  \mathcal{L}^{l}_{\mathrm{kin}}=i\overline{L_{L_i}}\gamma^{\mu}D_{\mu}L_{L_i}+ (D_{\mu}\Phi)^{\dagger}(D^{\mu}\Phi), \label{kinLi}\ee
where $D_{\mu}$  is the covariant derivative  defined in the SM.  All relevant couplings of the LFVHD are collected in  Table \ref{LFVCoupling}.
\begin{table}[t]
  \centering
\begin{tabular}{|c|c|c|c|}
  \hline
  Vertex & Coupling & Vertex & Coupling \\
   \hline
  $h\overline{e_i}e_i$ &$ -\frac{i m_{e_i}}{v}c_{\alpha}$  &  $h\overline{\nu_i}\nu_i$ &$ \frac{i m_{\nu_i}}{v'}s_{\alpha}$  \\
  \hline
  $h\overline{N_i}N_i$ &$ \frac{i m_{N_i}}{v'}s_{\alpha}$  &  $hW^+_{\mu}W^-_{\nu}$ &$i gm_Wc_{\alpha}g^{\mu\nu}$ \\
  \hline
  $h h^+_1h^-_1$ &$ i\left(-vc_{\alpha}\lambda_{\Phi h_1}+ v' s_{\alpha}\lambda_{\Sigma h_1}\right)$  &   $h h^+_2h^-_2$ &$ i\left(-vc_{\alpha}\lambda_{\Phi h_2}+ v's_{\alpha}\lambda_{\Sigma h_2}\right)$   \\
  \hline
 $h(p_0)W^+_{\mu}G_w^-(p_-)$ &$\frac{i g}{2}c_{\alpha}(p_0-p_-)^{\mu}$ &  $h(p_0)W^-_{\mu}G_w^+(p_+)$ &$-\frac{i g}{2}c_{\alpha}(p_0-p_+)^{\mu}$    \\
  \hline
  $\overline{e_i}\nu_jh^-_1$ &$-i(y^T_LU^L)_{ij}P_R$ &   $\overline{\nu_j}e_ih^+_1$ &$i(y^T_LU^{L})^*_{ij}P_L$  \\
  \hline
  $\overline{e_i}\nu_jG^-_w$ &$-i\frac{\sqrt{2}m_{e_i}}{v}U^L_{ij}P_L$ &   $\overline{\nu_j}e_iG^+_w$ &$-i\frac{\sqrt{2}m_{e_i}}{v}U^{L*}_{ij}P_R$  \\
  \hline
   $\overline{e_i}N_jh^-_2$ &$-i(y^T_R)_{ij}P_L$ &   $\overline{N_j}e_ih^+_2$ &$-i(y^T_R)_{ij}P_R$  \\
  \hline
  $\overline{e_i}\nu_j W^-_{\mu}$ &$\frac{ig}{\sqrt{2}}U^L_{ij}\gamma^{\mu} P_L$ &   $\overline{\nu_j}e_i W^+_{\mu}$ &$\frac{ig}{\sqrt{2}}U^{L*}_{ij}\gamma^{\mu} P_L$   \\
  \hline
   $h G^+_wG^-_w$ &$ i\left(-2vc_{\alpha}\lambda_{\Phi}+ v' s_{\alpha}\lambda_{\Phi\Sigma }\right)$   &  &  \\
   \hline
\end{tabular}
  \caption{Couplings of LFVHD in the t' Hooft-Feynman gauge. The momenta are incoming}\label{LFVCoupling}
\end{table}
\subsection{Parameter constraints from the previous work} 
For calculating the BR of the LFVHD, in the following sections we will mainly use the constraints of parameters obtained in \cite{Kenji}.  The important points are reviewed as follows. The parameters in the model were first investigated to ensure that they  satisfy the neutrino oscillation data, the current bounds of the BR of the LFV processes, the universality of charged currents, and  the vacuum stability of the Higgs self-couplings.  In addition, the doubly charged Higgses $k^{\pm\pm}$ are assumed to be light enough that they could be detected at the LHC. The constraints on parameters involving with the LFVHD are (i) the Dirac phase of the active  neutrino mixing matrix prefers the value of $\delta=\pi$, while the Majorana phase is still free;( ii) the masses of singly charged Higgs bosons should not be smaller than 3 TeV; (iii) the value of $|(y_R)_{22}|$ should be around 9; (iv) the value of $v'$ should be on the order of $O(1)$ TeV. The investigation in \cite{Kenji} also showed that the heavier doubly charged  Higgs bosons $k^{\pm\pm}$  will allow the lighter singly charged Higgs bosons. This leads to an interesting consequence of large values of the BR of the LFVHD as we will show in the numerical investigation.

The constraint from the LHC Higgs boson search was also discussed in \cite{Kenji}, including the effects of the $U(1)$ global Goldstone boson in the invisible decay of the SM-like Higgs bosons and the pair annihilation of the dark matter (DM) candidate $N_{R_1}$. From this, the constraint of mixing angle of neutral Higgs bosons is obtained as $|\sin\alpha|\leq 0.3$. Finally, the condition of DM candidate  mentioned above leads to the conclusion that the $N_{R_1}$ mass  should be around the value of $m_h/2$ in order to successfully explain the current relic density of DM; the VEV $v'$ was found smaller than $10$ TeV.

Many other issues involving with global $U(1)$ Goldstone boson were discussed in detail in \cite{Kenji}, for instance, anomaly-induced interaction to two photons, active-sterile neutral lepton mixing, and neutrinoless double beta decay via $W$ exchange.  Possible bounds from cosmological issue, such as the effect on cosmic microwave background via cosmic string generated by the spontaneous breaking down of the global $U(1)$ symmetry, were also mentioned. None of these issues  change  the constraints of parameters indicated above. Though new constraints on DM masses in the presence of $U(1)$ global symmetry were addressed in \cite{Gu1}, more studies are needed for confirmation. Furthermore, the considered global symmetry can be moved straight to the local one \cite{Kenji}, or replaced with a suitable discrete symmetry. This discussion is beyond the scope of this work.

In the next section, we will focus on parameters affecting the LFVHD and will discuss more clearly the relevant constraints if ones are needed.

\section{\label{para}Formulas of LFVHD and parameter constraints}
 The effective Lagrangian of  the decay  $h\rightarrow \tau^{\pm}\mu^{\mp}$  is  written as
$ \mathcal{L}^{LFV}= h \left(\Delta_L \overline{\mu}P_L \tau +\Delta_R \overline{\mu}P_R \tau\right) + \mathrm{h.c.}$,
  where   $\Delta_{L,R}$ are scalar factors arising from the loop contributions.
  The decay amplitude is defined as $i\mathcal{M}= i\bar{u}_{1}\left(\Delta_LP_L+\Delta_RP_R\right)v_{2}$ \cite{iseesaw},
 where $u_1\equiv u_1(p_1,s_1)$ and $v_2\equiv v_2(p_2,s_2)$ are the Dirac spinors of a muon and a tauon, respectively.
The partial width of  the decay is given as
\be
\Gamma (h\rightarrow \mu\tau)\equiv\Gamma (h\rightarrow \mu^{-} \tau^{+})+\Gamma (h\rightarrow \mu^{+} \tau^{-})
=  \fr{ m_{h} }{8\pi }\left(\vert \Delta_L\vert^2+\vert \Delta_R\vert^2\right), \label{LFVwidth}
\ee
 with the condition  $m_{h}\gg m_1,m_2$, where $m_1,m_2$ are muon and tauon masses, respectively. The on-shell conditions for external particles are $p^2_{i}=m_i^2$ (i=1,2) and $ p_h^2 \equiv( p_1+p_2)^2=m^2_{h}$.

 The  loop contributions can be separated  into two parts  $  \Delta_L= \Delta^{\nu}_L+ \Delta^{N}_L$ and $  \Delta_R= \Delta^{\nu}_R+ \Delta^{N}_R$, corresponding to the appearance of the active and exotic neutrinos in the loops. In the 't Hooft-Feynman gauge,  the specific formulas of contributions from diagrams shown in  Fig. \ref{LFVdiagram} are listed in  Appendix \ref{ffactor}, where new notations such as $E_{L,R}$ factors are used.  The contribution of the loops with active neutrinos $\Delta^\nu_L$ is obtained as
\bea  \Delta^{\nu}_L
&=& \frac{1}{16\pi^2} \sum_{a=1}^3  (y^T_LU^L)_{2a}(y_L^TU^L)^{*}_{3a}\left[ ( v' \lambda_{hh_1h_1}) E^{ \nu h_1h_1}_L
+s_{\alpha} E^{h_1\nu\nu }_L
+\left(-c_{\alpha}\right)E^{h_1\nu }_L \right]\crn
&+&\frac{1}{16\pi^2}\sum_{a=1}^3U^L_{2a} U^{L*}_{3a}\left[\frac{}{}
g^3c_{\alpha} E^{\nu WW}_L
+ \frac{g^2c_{\alpha}}{2} \left(E^{\nu G_wW}_L+ E^{\nu W G_w}_L\right)
\right.\crn  &+&
 \left(v' \lambda_{h G_w Gw}\right)E^{\nu G_w Gw}_L+
s_{\alpha} \left(\frac{g^2}{2}E^{W\nu\nu }_L+ E^{G_w\nu\nu }_L \right)
\left.-c_{\alpha}\left(\frac{g^2}{2} E^{W\nu }_L+ E^{G_w\nu }_L\right) \right],
 \label{DhnuL} \eea
where  $( v' \lambda_{hh_1h_1})=-vc_{\alpha}\lambda_{\Phi h_1}+v's_{\alpha}\lambda_{\Sigma h_1} $ and $\left(v' \lambda_{h G_w Gw}\right)= -2vc_{\alpha}\lambda_{\Phi }+v's_{\alpha}\lambda_{\Phi\Sigma } $ .

 The contribution from exotic neutrino mediations $\Delta^N_L$ is given as
\be  \Delta^{N}_L
= \frac{1}{16\pi^2}\sum_{a=1}^3 (y^T_R)_{2a} (y^T_R)_{3a}\left[ \left(v'\lambda_{hh_2h_2}\right) E^{N h_2h_2}_L
 + s_{\alpha} E^{h_2NN}_L
 -c_{\alpha}   E^{N h_2}_L\right],
\label{DhEnuL} \ee
where $\left(v'\lambda_{hh_2h_2}\right)=-vc_{\alpha}\lambda_{\Phi h_2}+v's_{\alpha}\lambda_{\Sigma h_2} $.

Similarly, we have  $\Delta^{\nu}_{R}= \Delta^{\nu}_L\left( E_L\rightarrow E_R\right)$ and   $\Delta^{N}_{R}= \Delta^{N}_L\left(E_L\rightarrow  E_R\right)$.

As proved in  Appendix \ref{ffactor},  the $\Delta_{L,R}$ are convergent.  Specifically, in the 't Hooft-Feynman gauge, the private contributions from specific diagrams  are  always finite.  In addition, the limit $p_1^2,p_2^2\simeq 0$ result in the extremely small contributions of  diagrams relating with the two point functions, namely \ref{LFVdiagram}g), 1h), 1i) and 1k).  Hence their contributions are ignored. Besides,  it can be estimated that  the sum of  contributions in the two last lines  in (\ref{DhnuL}) is very suppressed, because of the GIM mechanism, controlled   by the factor $\sum_{a=1}^3U^L_{2a} U^{L*}_{3a}$. This is a general property of all models where the neutrino masses are generated from the seesaw mechanism. Whereas  the contributions  in the first lines of (\ref{DhnuL}) and (\ref{DhEnuL}) may be large because the appearance of the $y_L$ and $y_R$ breaks the GIM mechanism,   non-zero contributions survive which do not contain factors of very light of neutrino masses.

In the numerical calculation, the following parameters are taken from  experimental data, for example \cite{PDG2014}:  $v\simeq 246$ GeV, $m_h=125.1$ GeV, $m_W=80.4$ GeV, and the muon and tauon masses are $m_{\mu}=0.105$ GeV, $m_{\tau}=1.776$ GeV. The total decay width of the SM-like Higgs bosons $\Gamma_h=4.1\times 10^{-3}$ GeV is used. Based on the investigation of  \cite{Kenji}, relevant parameters of the active neutrino masses are only considered in the normal hierarchy scheme. In particular the mixing parameters $U^L$ are expressed as follows
  \bea U^L=\left(
           \begin{array}{ccc}
             1 & 0 & 0 \\
             0 &c_{23} & s_{23} \\
             0 & -s_{23} & c_{23} \\
           \end{array}
         \right)
\left(
           \begin{array}{ccc}
            c_{13} & 0 & -s_{13} \\
             0 & 1 & 0 \\
            s_{13} & 0 & c_{13} \\
           \end{array}
         \right)
\left(
           \begin{array}{ccc}
            c_{12} & s_{12} & 0 \\
             -s_{12} & c_{12} & 0 \\
             0 & 0 & 1 \\
           \end{array}
         \right),\label{mixingpar}\eea
         where $c_{ij}\equiv \cos\theta_{ij}, s_{ij}\equiv \sin\theta_{ij}$ and the Dirac CP phase $\delta$ and the Majorana
       CP phase $\phi$ are taken as $ \delta=\pi$ and $\phi=0$.
  The best-fit  values of  neutrino oscillation parameters given in \cite{actnuUpdate}
\bea \Delta m^2_{21}&=& 7.50\times 10^{-5}\;\mathrm{ eV^2},\hs  \Delta m^2_{31}= 2.457\times 10^{-3}\; \mathrm{eV^2},\crn
s^2_{12}&=&0.304,\; s^2_{23}=0.452,\; s^2_{13}=0.0218 \label{nuosc}\eea
are also used in our numerical calculations.
 The lightest neutrino mass will be chosen as $ m_{\nu_1}= 10^{-10}$ GeV, satisfying the condition $\sum_{b}m_{\nu_b}\leq 0.5$ eV obtained from the cosmological constraint. The remain two neutrino masses are $m^2_{\nu_b}=m^2_{\nu_1}+\Delta m^2_{\nu_{b1}}$ where $b=2,3$.

The  other unknown parameters involving the LFVHD are: the VEV $v'$;  the mixing angle of the two neutral Higgs bosons $\alpha$; the exotic neutrino masses $m_{N_a}$; the new  Higgs masses $m_{h^\pm_1}, m_{h^\pm_2}$,  $m_H$;  the Yukawa coupling matrices $y_L$ and $y_R$;  and the trilinear Higgs self-couplings  $\lambda_{\Phi h_1},\lambda_{\Phi h_2}, \lambda_{\Sigma h_1},\lambda_{\Sigma h_2}$.

 In the numerical investigation,  we focus first on the most interesting regions of the parameter space  indicated in \cite{Kenji}, where $m_{N_1}$ plays the role of a DM  particle and doubly charged Higgses $k^{\pm\pm}$ are light enough to be observed at the LHC. The values of the relevant parameters are summarized as follows:    $m_h/2=m_{N_1}<m_{N_2}<m_{N_3}$, $|s_{\alpha}|\leq 0.3$; 3 TeV$\leq (m_{h^\pm_1},m_{h^\pm_2})\sim \mathcal{O}(1)$ TeV and $v'\sim \mathcal{O}(1)$ TeV.  The Yukawa coupling matrix $y_L$ are satisfied the following conditions \cite{yLbound},
 \bea (y_L)_{13} &=&\left( \frac{s_{12}c_{23}}{c_{12}c_{13}}- \frac{s_{13}s_{23}}{c_{13}} \right) (y_L)_{23}=0.394 (y_L)_{23},\crn
   (y_L)_{12} &=& \left( \frac{s_{12}s_{23}}{c_{12}c_{13}}+ \frac{s_{13}c_{23}}{c_{13}} \right) (y_L)_{23}=0.56(y_L)_{23} ,\label{yLbound}\eea
 and $|(y_L)_{23}|\leq 1$. The conditions of gauge coupling universalities also imply that
 \be |(y_L)_{23}|< \frac{0.3 m_{h^{\pm}_1}}{1\mathrm{TeV}}. \label{Gun}\ee
The most stringent constraint comes from the LFV decay of muon with Br$(\mu\rightarrow e\gamma)<5.7\times 10^{-13}$. It gives the direct upper bounds on the following products of  the Yukawa couplings: (i) $(y^T_L)_{23}(y^T_L)_{13}$ in the loop including virtual active neutrinos and $h^{\pm}_1$ since $y_L$ is antisymmetric; (ii) $(y^T_R)_{2i}(y^T_R)_{1i}$ ($i=1,2,3$) in the loops including exotic neutrinos and $h^{\pm}_2$. The other constraints from the tauon decays are less stringent, hence are omitted  here.  On the other hand, the  Br$(h\rightarrow \mu\tau)$ depends on the products $(y^T_R)_{2i}(y^T_R)_{3i} $ with $i=1,2,3$. So, if $|(y^T_R)_{ij}|$ with $(ij)=\{(11), (21), (31)\}$ is small enough, the values of $ (y^T_R)_{ij}$ with $(ij)=\{(22),(32), (33)\}$ may be large, without any inconsistence in the upper bounds of the BR in the LFV decays of charged leptons. In order to find the reasonable regions of parameter space,  the upper bounds must be checked in the formula given in \cite{Kenji}:
\be \frac{\sum_{a=1}^3\left[ (y^{\dagger}_L)_{af}  (y_L)_{i a}\right]^2(I_{1,a}I_{2,a}+I^2_{1,a}) + \sum_{a=1}^3\left[ (y_R)_{af}  (y^{\dagger}_R)_{ia}\right]^2(I'_{1,a}I'_{2,a}+I'^2_{1,a})  }{16m^4_{h^\pm_1}\left|\sum_{a=1}^3(I_{1,a}I_{2,a}+I^2_{1,a}) \right|}< \frac{C_{if}}{[\mathrm{TeV}]^4}, \label{muegamma} \ee
where $(i,f)=(\mu,e)=(2,1)$, $C_{if}=1.6\times 10^{-6}$ and many notations including $(I_{1,a},I_{2,a},I'_{1,a},I'_{2,a})$ are defined in detail in \cite{Kenji}. For simplicity, we mention only the following special cases.  In the limit $0\simeq m_{\nu_a}\ll m_{h^\pm_1}$ and $m_{h^\pm_2} \gg m_{N_1}$ we have
\be I_{1,a}\simeq -\frac{1}{(4\pi)^2}\frac{1}{36 m^2_{h^\pm_1}},\; I_{2,a}\simeq -\frac{1}{(4\pi)^2}\frac{5}{36 m^2_{h^\pm_1}},\;
 I'_{1,a}\simeq -\frac{1}{(4\pi)^2}\frac{1}{36 m^2_{h^\pm_2}},\; I'_{2,a}\simeq -\frac{1}{(4\pi)^2}\frac{5}{36 m^2_{h^\pm_2}}.\label{smalnu}\ee
  While a very large $m_{N_a}$ gives $I'_{1,a},I'_{2,a}\simeq 0$-for example when $m_{N_{2,3}}$ are  larger than few TeV- the bound  (\ref{muegamma})  affects only the  products $|(y^\dagger_L)_{31}(y_L)_{23}|$ and $|(y_R)_{31}(y_R)^\dagger_{23}|$. Combining this with (\ref{yLbound}) we get new constraints:
 \be \left|(y_L)_{23}\right|\leq 0.149 \frac{m_{h^\pm_1}}{1 \mathrm{TeV}},\hs  \mathrm{and}\; |(y_R)_{31}(y_R^\dagger)_{23}|< \left( \frac{0.1 m_{h^\pm_2}}{1 \mathrm{TeV}}\right)^2. \label{yLRbound} \ee
   This constraint of $|(y_L)_{23}|$  is consistent with the numerical investigation done in \cite{Kenji}, where $m_{h^\pm_1}=4.8$ TeV prefers $(y_L)_{23}=0.5-0.6<0.7\simeq 4.8\times 0.15$. Interestingly,  the first inequality in (\ref{yLRbound}) is more strict than the one given in (\ref{Gun}). The second constraint in (\ref{yLRbound}) suggests that  the small $|(y_R)_{31}|$ will allow  large $|(y_R^\dagger)_{23}|$, i.e  the choice $ |(y_R)_{31}| \leq 10^{-3} \times (m_{h^\pm_{2}}/1\mathrm{TeV})^2$ will allow $|(y_R^\dagger)_{23}|\sim \mathcal{O}(1)$.  The $m_{h^\pm_2}=5 (3)$ TeV will give $  |(y_R)_{31}| \leq 0.25(0.09)$, being  consistent with the promoting regions indicated in \cite{Kenji}.  But the absolute values of $(y_{R})_{22}$ and $(y_{R})_{23}$ should be smaller than the perturbative upper bound $ 4\pi$.

  Another relevant constraint  is the small upper bound of the BR$(\mu\rightarrow 3e)$.  Following \cite{Kenji},  all  form factors relating with $(y_{R})_{ij}$ contain at least three factors $(y_{R})_{i1}$ or $(y_{R})_{1i}$. Therefore they result in very suppressed values of the BR of the LFVHD if  all $|(y_R)_{i1,1i}|$'s are small enough, without  any conditions of small $|(y_R)_{23,32,33}|$'s.

Combining the above discussion with  the analysis in \cite{Kenji},  the  reasonable values of the free parameters can be chosen as follows: $s_{\alpha}=0.3,\lambda\equiv8\lambda_{\Phi h_1}= 8\lambda_{\Phi h_2}=\lambda_{\Sigma h_1}=\lambda_{\Sigma h_2}=4 $, $m_{h_1^{\pm}}=m_{h_2^{\pm}}=3 $ TeV, $m_H=2$ TeV, $(y_L)_{23}= \frac{0.14 m_{h_1^{\pm}}}{1\mathrm{TeV}}$, $y'_R=(y_R)_{23}=(y_R)_{32}=(y_R)_{33}=3,(y_R)_{22}=8$, $v'\in\{1,2,8,10\}$ TeV, $m_{N_1}=m_h/2, m_{N_2}=1/2m_{N_3}=5$ TeV and $(y_R)_{i,j}=10^{-2}$ with at least one of the indices, $i$ or $j$ being 1.  We would like to stress that the above choices are also based  on the following additional reasons. The  values of $(y_L)_{23}$ and $(y_{R})_{ij}$ always satisfy all recent bounds of the BR of the LFV decays of charged leptons (\ref{yLRbound}) as well as gauge coupling universalities (\ref{Gun}).
 The $\lambda$ parameter is positive and small enough  to satisfy both conditions of  perturbative limit and vacuum stability, whereas it  is  large enough to  enhance the  BR  of LFVHD.

 To investigate the variance of BR$(h\rightarrow \mu\tau)$ versus the changing of free parameters,  the ranges of free parameters will be chosen as follows: $|s_{\alpha}|\leq 0.4$, $0.1<\lambda<10$, $|y'_R|\leq4$, $|(y_L)_{23}|\leq \frac{0.15 m_{h^\pm_1}}{1\mathrm{TeV}}$, $1\mathrm{TeV} \leq m_{h^\pm_1},m_{h^\pm_2}, v'<10$ TeV and $0.5\mathrm{ TeV}\leq m_{N_2}\leq 6$ TeV.

\section{\label{numerical}Numerical result and discussion}

In this section, we will first investigate some private contributions to the BR of the LFVHD, namely the active neutrino loops with $W^\pm$/ $h_1^{\pm}$ bosons, and the exotic lepton loops. Based on this,   the parameter space regions which give the large total contribution will be further studied.

\begin{figure}[h]
  \centering
 \begin{tabular}{cc}
  \includegraphics[width=8cm]{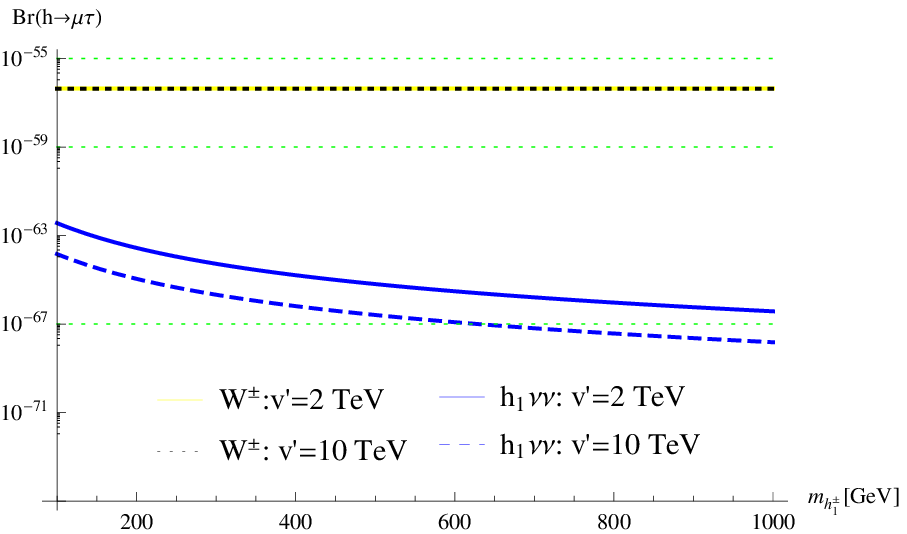}& \includegraphics[width=8cm]{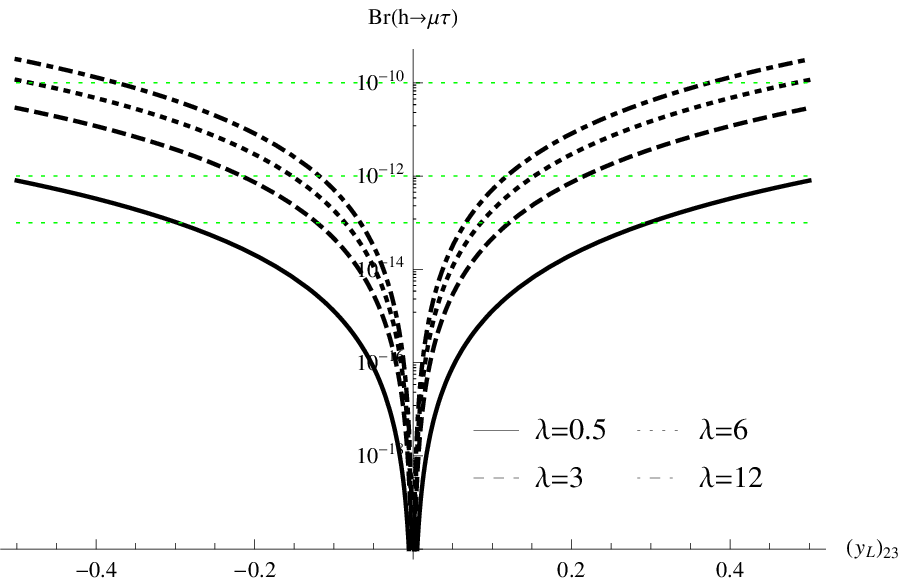}\\
  \end{tabular}
  \caption{Private contributions of only active neutrino loops to the Br($h\rightarrow \mu\tau$), where the left hand side is assumed that $(y_L)_{23}=\frac{0.14 m_{h^\pm_1}}{1\mathrm{TeV}}$ and the right hand side corresponds to $v'=10$ TeV. }\label{nucontrution}
\end{figure}

The contributions of active neutrinos are shown in the Fig. \ref{nucontrution}. The left panel shows two contributions to the Br($h\rightarrow \mu\tau$):  sum of all diagrams relating with the $W^{\pm}$ gauge bosons and their Goldstone bosons, and Fig. \ref{LFVdiagram}f). The right panel shows the contribution to Br($h\rightarrow \mu\tau$) from Fig.  \ref{LFVdiagram}d) with virtual $h^{\pm}_1$ in the loop.
The figure emphasizes  two points: the tiny contributions in the left panel and the significantly enhancement of the contribution in the right panel.  Similarly to many well-known models, the total contribution of electroweak loops, including $W^\pm$ and their Goldstone bosons, is very suppressed because of the GIM mechanism, arising  from the sum of two different flavors of external lepton fields: $\sum_aU^L_{2a}U^{L*}_{3a}=0$. This sum cancels the largest terms of the contributions when they are expanded in terms of $(m_{\nu_a}/m_W)^2$ series. Only terms containing factors $(m_{\nu_a}/m_W)^2$ survive but they are very suppressed.  In addition, this contribution does not depend on the $v'$, leading to the overlap lines in the left panel of the figure. The second contribution in the left panel comes from the $\nu\nu h_1^{\pm}$ loops. Although the appearance of the Yukawa couplings $(y_L)_{ij}$ removes the GIM mechanism,  the contribution its self contains a factor of $m^2_{\nu_a}$; therefore it is suppressed, too. It is even  smaller than the electroweak-loop contribution because $m_{h^{\pm}_1}$ is much larger than $m_W$. The $\nu h^\pm_{1}h^\pm_{1}$ is much enhanced because of the presence of both large coupling $\lambda_{hh^\pm_1h^\pm_1}$ and $y_L$.   In the  model considered, $|(y_{L})_{23}|$ is much constrained from (\ref{yLRbound}), where the $m_{h^\pm_1}=3$ TeV gives the small $|(y_L)_{23}|<0.45$. Also, the $\lambda_{hh^\pm_1h^\pm_1}$, as functions of $v'~ (<10 \mathrm{TeV})$ and Higgs self-couplings ($<4\pi$), does not allow large values of  Br($h\rightarrow \mu\tau$). Then the contribution of the $\nu\nu h^{\pm}_1$ loop to the Br($h\rightarrow \mu\tau$) is not larger than $10^{-10}$. In any case, this provides a hint for enhancing the contribution from active neutrino loops, for example in models with a four-loop (or higher) neutrino mass and small $m_{h^\pm_1}$, where the constraint of $y_L$ may be released. This deserves  further study.

\begin{figure}[h]
  \centering
 \begin{tabular}{cc}
  \includegraphics[width=8cm]{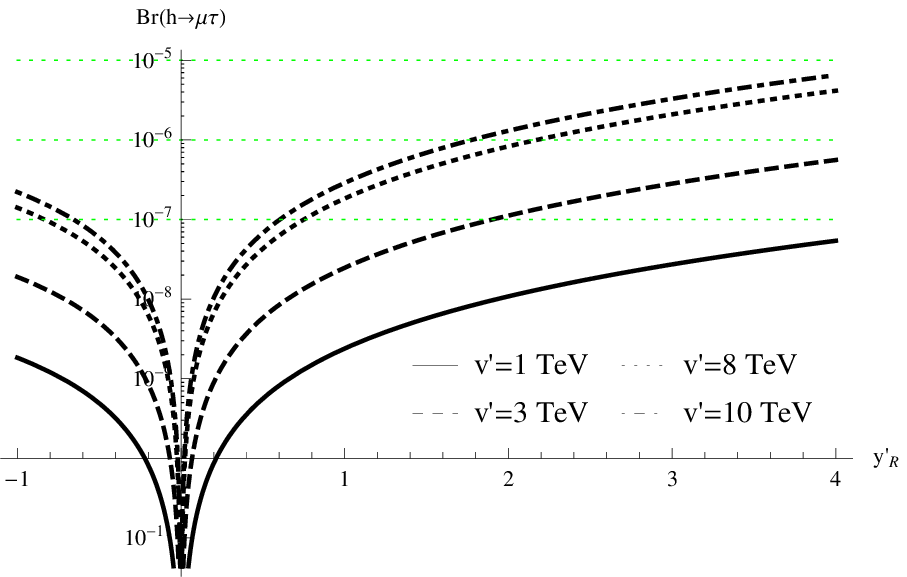}& \includegraphics[width=8cm]{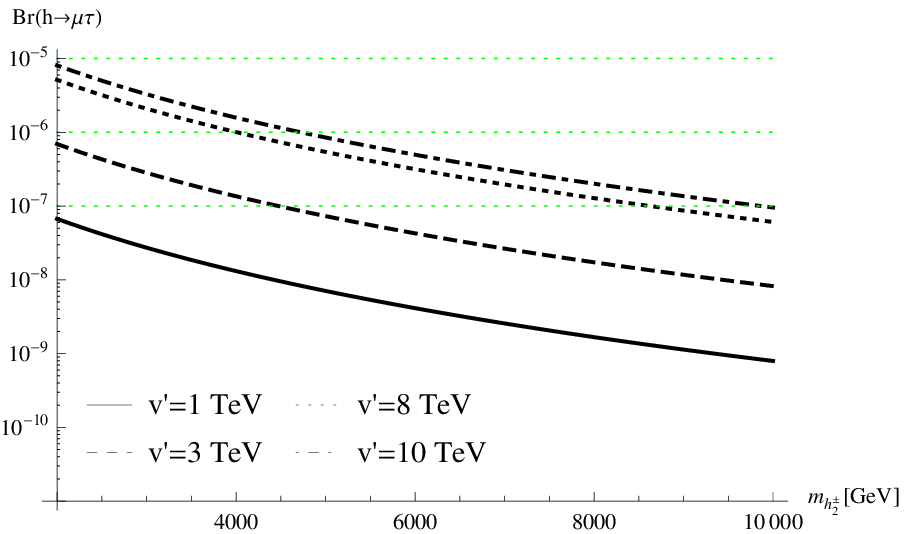}\\
  \end{tabular}
  \caption{Private contributions of only $Nh^{\pm}_2h^{\pm}_2$ loops, i.e.,  diagrams  \ref{LFVdiagram} d),  to the Br($h\rightarrow \mu\tau$) as functions of $y'_{R}$ ($m_{h_2}$) in the left  (right) panel.}\label{cNh2h2}
\end{figure}

\begin{figure}[h]
  \centering
 \begin{tabular}{cc}
  \includegraphics[width=8cm]{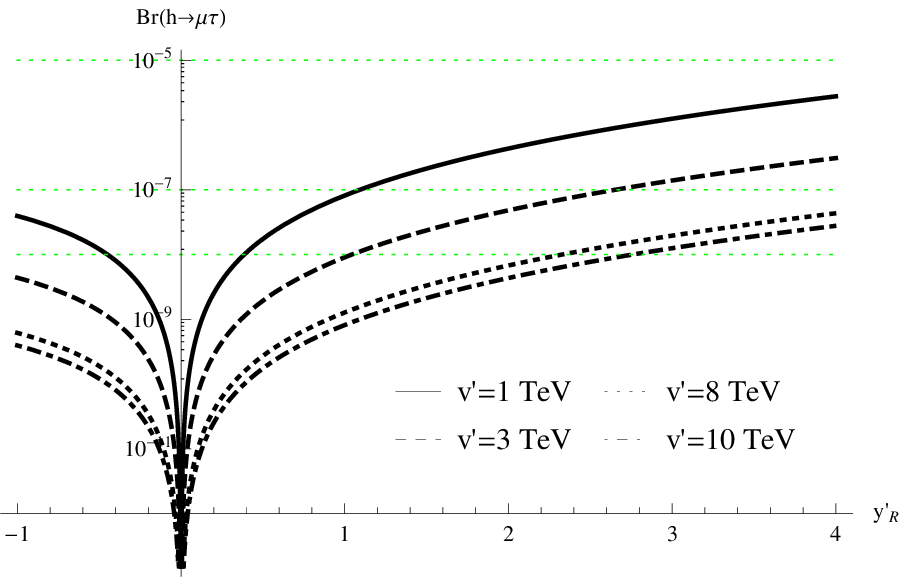}& \includegraphics[width=8cm]{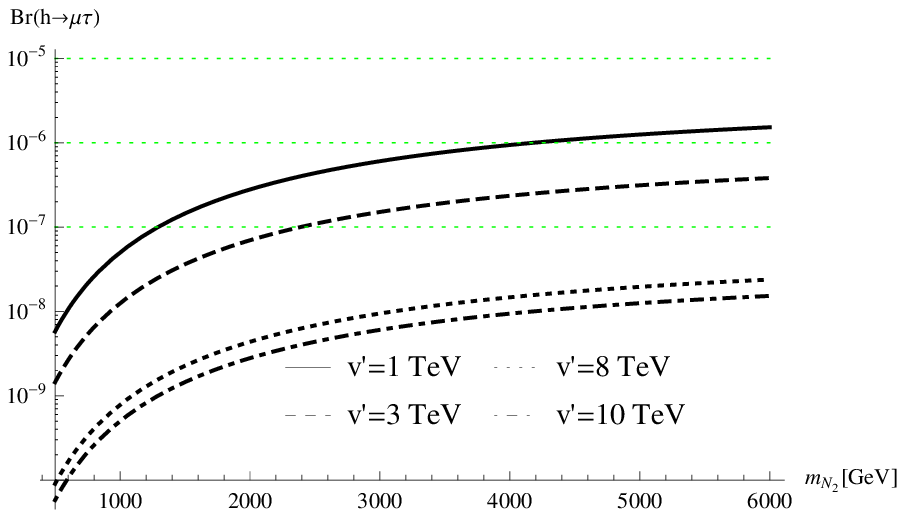}\\
  \end{tabular}
  \caption{Private contributions of only $h^{\pm}_2NN$ loops, i.e.,  diagrams  \ref{LFVdiagram} f),  to the Br($h\rightarrow \mu\tau$) as functions of $y'_{R}$ ($m_{N_2}$) in the left  (right) panel.}\label{ch2NN}
\end{figure}

The main contributions of exotic leptons to the Br($h\rightarrow \mu\tau$) come from the two diagrams $Nh^\pm_2h^\pm_2$  and $h^\pm_2NN$.  They do not depend on $y_L$ but depend strongly on $|(y_R)_{ij}|$ with $\{i,j\}=\{2,3\}$.  As illustrated in Figs. \ref{cNh2h2} and \ref{ch2NN}, the two contributions have common properties. They increase with a decreasing $m_{h^\pm_2}$, in the same behaviour shown in the right panel of the Fig. \ref{cNh2h2}.  Most important is that they are enhanced strongly with an increasing $|y'_R|$,  see  the two left panels of the two figures.

On the other hand, these two contributions behave in opposing ways with the changes of $v'$ and $m_{N_2}(m_{N_3})$.  The contribution from $m_{N_1}$ mediation is ignored because of a very small $|(y_R)_{i1,1i}|$ ($i=1,2,3$).  The $Nh^\pm_2h^\pm_2$ mediation relates with the Higgs-self coupling; hence its contribution is large with a small $m_{N_2}$ and a large $v'$. See the figure \ref{cNh2h2} with its four fixed values of $v'$, where the largest corresponds to $v'=10$ TeV.  For the $h^\pm_2NN$ loops, their analytic expression contains   $m_{N_a}^2/v'$ factors separate from the $C$ functions. Therefore, the small $v'$ and large $m_{N_a}$ will give large contributions; see in the right panel of  Fig. \ref{ch2NN}.

\begin{figure}[h]
  \centering
 \begin{tabular}{cc}
  \includegraphics[width=8cm]{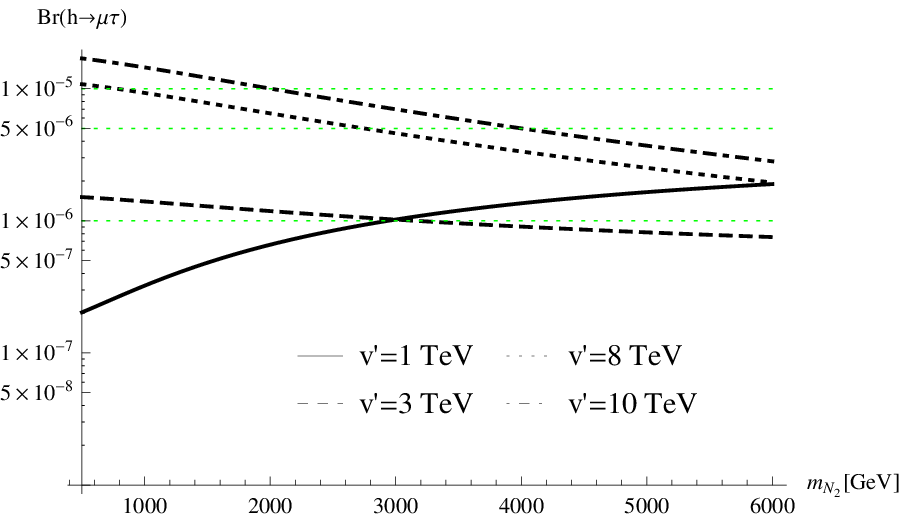}& \includegraphics[width=8cm]{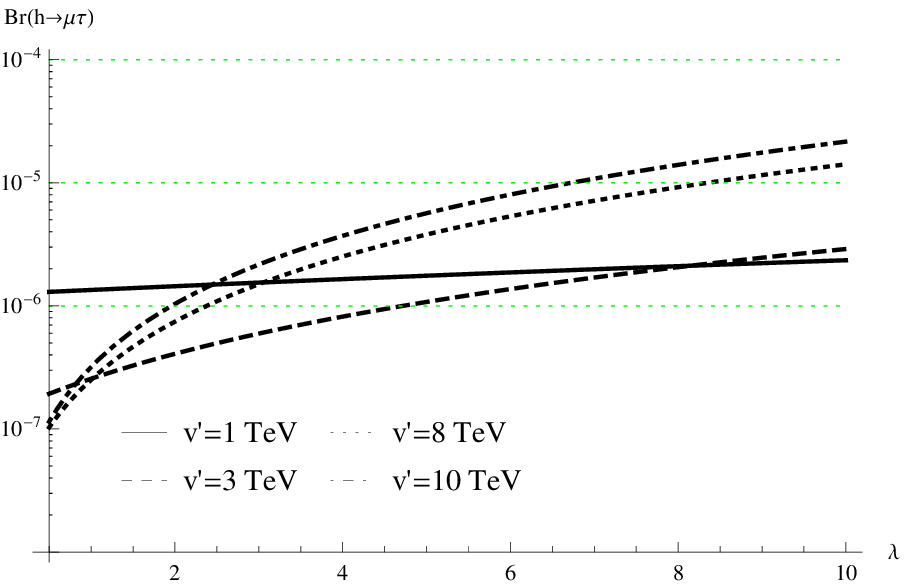}\\
  \includegraphics[width=8cm]{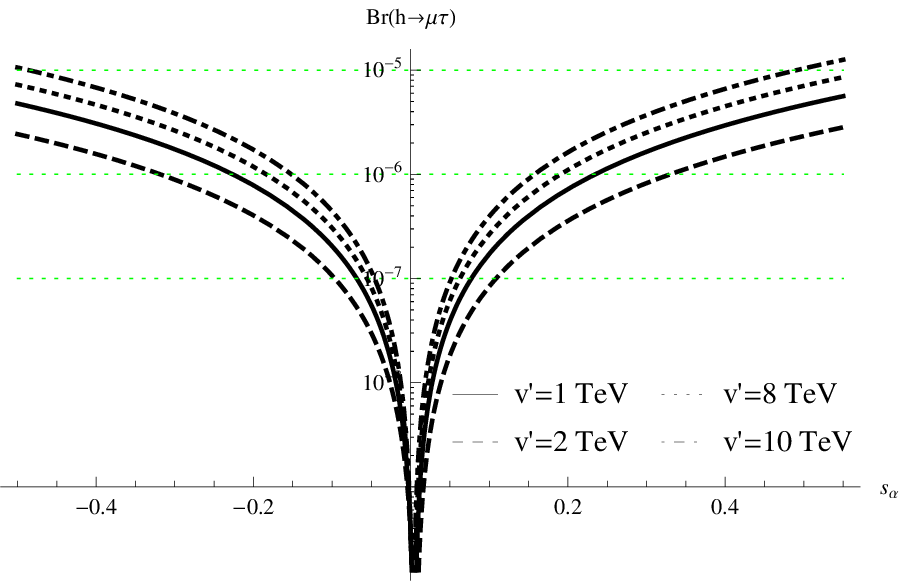}& \includegraphics[width=8cm]{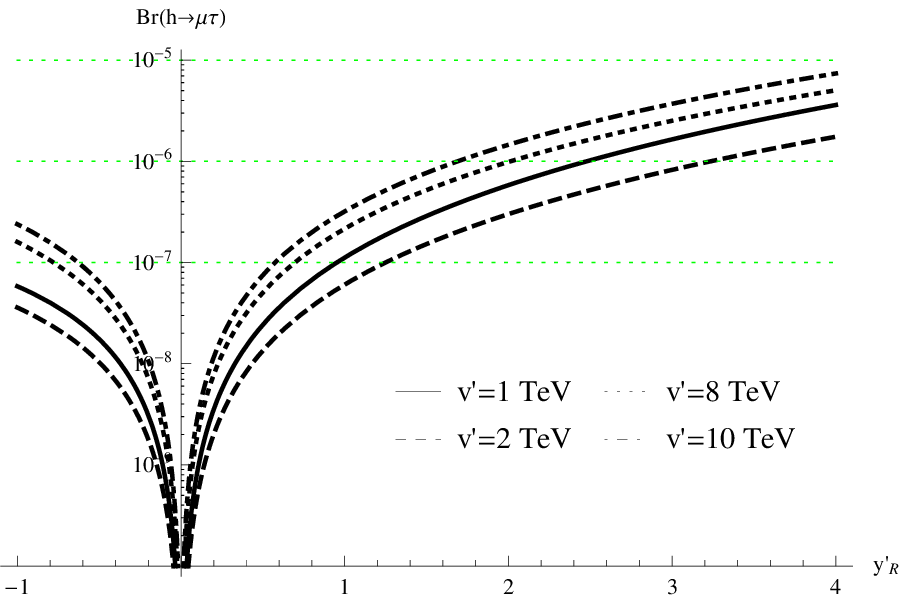}\\

  \end{tabular}
  \caption{The total contribution  to the Br($h\rightarrow \mu\tau$) as functions of single parameters: $m_{N_2}$, $\lambda$, $s_{\alpha}$ or $y_R'$.}\label{ftot}
\end{figure}

All of the above discussions suggest that the Br$(h\rightarrow\mu\tau)$ will be large with small singly charged Higgs masses and large values of all of the following parameters: the coupling $|y'_R|$, $|(y_L)_{23}|$ and Higgs-self coupling $\lambda$. The dependence of the BR on $v'$ and $m_{N_{2,3}}$ is  a bit complicated. The dependence on the mixing angle $\alpha$ of two CP-even neutral Higgses should also be mentioned. Figure \ref{ftot} shows more precisely the variations of the Br$(h\rightarrow \mu\tau)$ on some particular free parameters. We can realize that $y'_R$ affects  the change of this BR the most significantly.  It is larger than $10^{-6}$ only when $|y'_R|\geq2$.  Br($h\rightarrow \mu\tau$)  does not depend  on the signs of  $s_{\alpha}$ and $y'_R$, but it does depend significantly on the absolute values of these parameters.  With $\lambda = 4$, $m_{h^\pm_2}=3$ TeV and $s_{\alpha}=0.3$ , the Br($h\rightarrow \mu\tau$) can reach $10^{-5}$ when all of these conditions are satisfied: $|y'_R|\geq3$,  $ v'\geq 8 $ TeV and $m_{N_{2,3}}$ are small enough.

\begin{figure}[h]
  \centering
  \begin{tabular}{cc}
  \includegraphics[width=8cm]{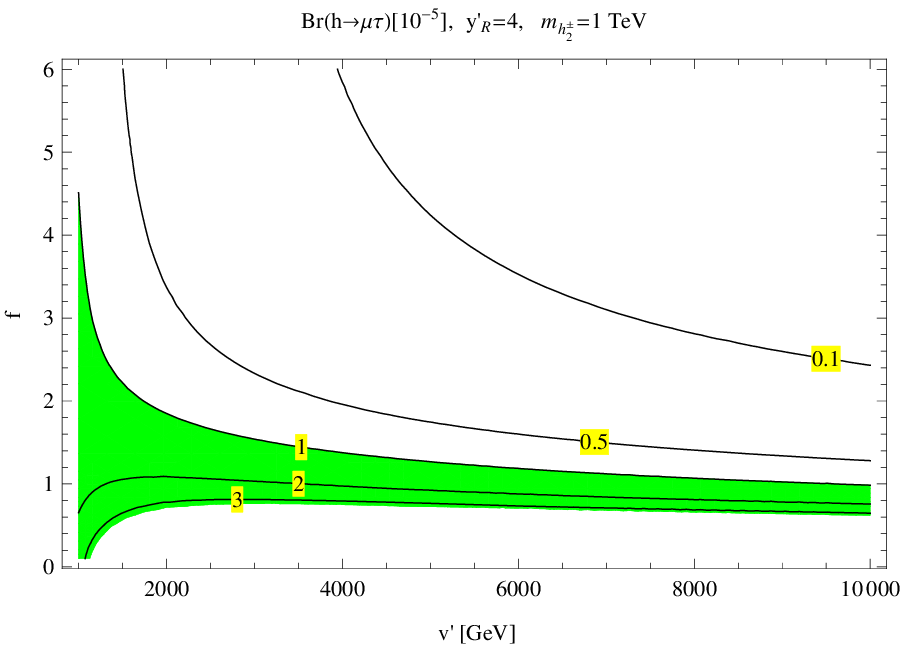}& \includegraphics[width=8cm]{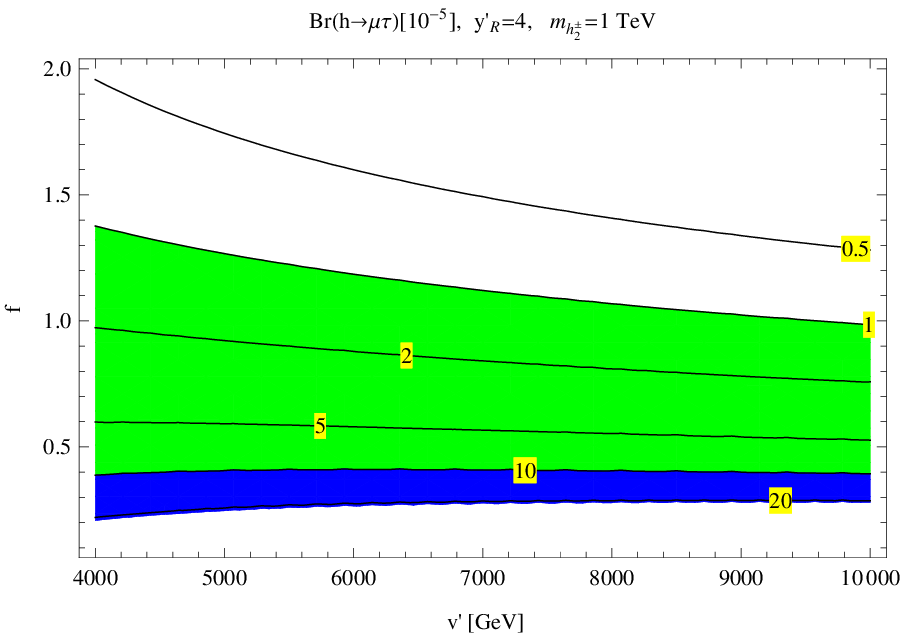}\\
  \end{tabular}
 \caption{ Contour plots  of Br$(h\rightarrow\mu\tau)$ as functions of  $v'$ and $f=m_{N_2}/v'$. The green and blue regions respectively  satisfy $10^{-5}\leq \mathrm{BR}(h\rightarrow\mu\tau)\leq 10^{-4}$  and $\mathrm{ BR}(h\rightarrow\mu\tau) \geq 10^{-4}$. }\label{conLFVH}
\end{figure}

Finally, if  the doubly charged Higgs bosons are heavy enough, the investigation shown in \cite{Kenji} may allow the presence of light, singly charged Higgs bosons $h^\pm_2$, for instance with the mass of 1 TeV.  Besides, if we define  $m_{N_2}=m_{N_3}/2=f\times v'$,  adopting also that $s_{\alpha}=0.3$ gives $v'\leq 9$ TeV \cite{Kenji,Kane}.  In addition, if $y'_R=4$, then BR$(h\rightarrow\mu\tau)$ can reach the value of $10^{-4}$, very close to the value noted by CMS.  This conclusion is illustrated in Fig. \ref{conLFVH}, where the $f$ parameter is scanned in the range of $0.1\leq f \leq 6$ ($0.1\leq f \leq 2$) in the left (right) panel. From our numerical calculation, the values of  $y'_R=4$ and $m_{h^\pm_1}=1$ TeV are the smallest ones  to obtain values about $10^{-4}$  of Br($h\rightarrow \mu\tau$). Accordingly, this large value lies in the only region where  $v'$ is larger than 4 TeV, while  $m_{N_2}$ and $m_{N_3}$  should be as small as possible. In this calculation, $m_{N_2}\geq 400$ GeV and the $Nh^\pm_2h^\pm_2$ contribution is  dominant. With small $v'\simeq 1$ TeV, the value around $10^{-5}$ of the Br($h\rightarrow \mu\tau$) may occur  if $m_{N_{2,3}}$ are large enough: $m_{N_2}\geq 3 v'$.  And the Br($h\rightarrow \mu\tau$) is unchanged, with general condition $m_{N_2}\leq m_{N_3}$, instead of  the $M_{N_2}=1/2 M_{N_3}$ one used above. In this case, the large LFVHD  corresponds to the dominant contribution from $h^\pm_2 NN$ diagram. The heavy neutrinos may be detected in the future lepton colliders \cite{Nueecolider}.

The singly charged Higgses are now being searched by experiments \cite{cHseach,PDG2014}. However, in the model considered,  none of them  are their targets, because they only couple with leptons. In addition, the $h^\pm_2$ hold negative parities and hence cannot decay to only normal leptons. We also believe that the condition (\ref{yLRbound}) is enough to guarantee  for constraints of LFV processes without the need for singly charged Higgs bosons that are too heavy. Hence, the  mass around 1 TeV of $h^\pm_2$ is reasonable for stability of the lightest $N_{R_1}$ as a DM candidate.

\section{\label{Con}Conclusion}
Radiative neutrino mass models are interesting ones for explaining the tiny masses of active neutrinos. In the  model introduced in \cite{Kenji}, the new parameters that generate neutrino masses radiatively are strongly constrained from recent experimental data such as neutrino oscillations, rare decay of charged leptons, gauge coupling universalities and other constraint from LHC Higgs boson physics. Therefore these parameters are very predictive for other phenomenologies such as dark matter, LFVHD, etc... In this work, we have shown that in the allowed region with heavy, singly charged Higgs bosons, where their masses are around 3 TeV, the BR of the LFVHD can reach a value of $10^{-5}$. If the mass of the $h^\pm_2$ is around 1 TeV, the Br($h\rightarrow \mu\tau$) can reach $10^{-4}$. The additional necessary conditions are $v'$, $|s_{\alpha}|$ and all amplitudes of Yukawa and trilinear Higgs couplings must be large enough. For example, with  largest $ s_{\alpha}=0.3$  to satisfy the LHC Higgs boson constraint, $v'$ should be in the range of $8-9$ TeV. Also,  $\lambda_{\Sigma h_2}\geq 4$ and $|y'_R|\geq3 (4)$ for largest values $10^{-5}(10^{-4})$ of the Br($h\rightarrow \mu\tau$). The masses of the two heavy exotic neutrinos should be small,  around 400 GeV for the lighter $m_{N_2}$.  In the  model under consideration,  a large BR$(h\rightarrow\mu\tau)$  will lead to  necessary  consequences: the doubly charged Higgses $k^{\pm\pm}$ must be heavy and the  BR$(h\rightarrow e\tau)$ must be very small. The latter can be explained from the constraint of $\mu\rightarrow e\gamma$. This requires a large $y'_R\simeq \mathcal{O}(1)$ to give very small $|(y_R)_{13,31}|\leq 10^{-3}$, implying that BR$(h\rightarrow e\tau)\leq 10^{-10}$, much smaller than recent sensitivity of experiments \cite{LFVhtauemu}.  One more interesting property is that the contribution of virtual active neutrinos may be large if the upper bound (\ref{yLRbound}) is ignored.  When this bound is considered the private contributions of $\nu h^\pm_1h^\pm_1$ to Br$(h\rightarrow\mu\tau)$  are around $10^{-10}$, which is much larger than values predicted by canonical seesaw models and \cite{LFVHDUgauge}. In  models with more than three-loop neutrino mass such as \cite{m3loop}, the bound  (\ref{yLRbound}) may be released.  We then guess that the active neutrino contributions can reach $10^{-7}$ or higher, and hence they should not be ignored. Inclusion,  more precise predictions can be worked out after the updating of data from the near future experiments.

\section*{Acknowledgments}
  The authors thank Dr. Farinaldo Queiroz for his comments of DM. This research is funded by Vietnam
   National Foundation for Science and Technology Development (NAFOSTED) under grant number 103.01-2015.33.
\appendix
\section{\label{PVfunction}One loop Passarino-Veltman functions}
Calculation in this section relates to one-loop diagrams
 in Fig. \ref{LFVdiagram}.  The analytic expressions of the PV function are given in \cite{LFVHDUgauge} and the needed formulas will be summarized in this appendix.  We would like to stress that these PV functions were derived from the general form given in \cite{Hooft}, using only the conditions of very small masses of tauons and muons. They are consistent with \cite{bardin}. The denominators of the propagators  are denoted as  $D_0=k^2-M_0^2+i\delta$, $D_1=(k-p_1)^2-M_{1}^2+i\delta$ and $D_2=(k+p_2)^2-M_2^2+i\delta$, where $\delta$ is  infinitesimally a positive real quantity. The scalar integrals are defined as
 \bea
 B^{(1)}_0 &\equiv&\frac{\left(2\pi\mu\right)^{4-D}}{i\pi^2}\int \frac{d^D k}{D_0D_1},
 \hs  B^{(2)}_0\equiv \frac{\left(2\pi\mu\right)^{4-D}}{i\pi^2}\int \frac{d^D k}{D_0D_2},  \crn
  B^{(12)}_0 &\equiv& \frac{\left(2\pi\mu\right)^{4-D}}{i\pi^2}\int \frac{d^D k}{D_1D_2}, \hs
 C_0\equiv  C_{0}(M_0,M_1,M_2) =\frac{1}{i\pi^2}\int \frac{d^4 k}{D_0D_1D_2},
 \label{scalrInte}\eea
 where $i=1,2$.
   In addition, $D=4-2\epsilon \leq 4$ is the dimension of the integral;  $M_0,~M_1,~M_2$ are masses of virtual particles in the loop. The momenta satisfy conditions: $p^2_1=m^2_{1},~p^2_2=m^2_{2}$ and $(p_1+p_2)^2=m^2_{h}$. In this work, with $m_1$ and $m_2$ are respective masses of the muon and the tauon, and $m_h$ is the SM-like Higgs boson mass. The tensor integrals are
 \bea
 B^{\mu}(p_i;M_0,M_i)&=& \frac{\left(2\pi\mu\right)^{4-D}}{i\pi^2}\int \frac{d^D k\times
k^{\mu}}{D_0D_i}\equiv B^{(i)}_1p^{\mu}_i,\crn
 B^{\mu}(p_1,p_2;M_1,M_i)&=& \frac{\left(2\pi\mu\right)^{4-D}}{i\pi^2}\int \frac{d^D k\times
k^{\mu}}{D_1D_2}\equiv B^{(12)}_1p^{\mu}_1+B^{(12)}_2p^{\mu}_2,\crn
C^{\mu} &=&C^{\mu}(M_0,M_1,M_2)=\frac{1}{i\pi^2}\int \frac{d^4 k\times k^{\mu}}{D_0D_1D_2}\equiv  C_1 p_1^{\mu}+C_2 p_2^{\mu},
 \label{oneloopin1}\eea
where  $B^{(i)}_{0,1}$ and $C_{0,1,2}$   are PV functions.  It is well known that $C_{i}$ is finite while the remains are divergent. We define
$\Delta_{\epsilon}\equiv \frac{1}{\epsilon}+\ln4\pi-\gamma_E+\ln\frac{\mu^2}{m_h^2}$ where $\gamma_E$ is the  Euler constant.  The divergent parts of the above scalar factors can be determined as
\be   \mathrm{Div}[B^{(i)}_0]= \mathrm{Div}[B^{(12)}_0]= \Delta_{\epsilon},
\mathrm{Div}[B^{(1)}_1]=- \mathrm{Div}[B^{(2)}_1]= \frac{1}{2}\Delta_{\epsilon}. \label{divs1}\ee
The  finite parts  of  the PV-functions such as B-functions  depend  on the scale of $\mu$ parameter with the same coefficient of the divergent parts.

 The analytic formulas of the above PV-functions are 
 \be B^{(i)}_{0,1}= \mathrm{Div}[B^{(i)}_{0,1}]+ b^{(i)}_{0,1}, \hs   B^{(12)}_{0,1,2}= \mathrm{Div}[B^{(12)}_{0,1,2}]+ b^{(12)}_{0,1,2}. \label{B01i}\ee
The  expression of $b^{(12)}_0$ is
 \be  b_0^{(12)}=\ln \frac{m_h^2-i\delta}{M_1^2-i\delta}+2 + \sum_{k=1}^2 x_k\ln\left(1-\frac{1}{x_k}\right),  \label{b012f1}\ee
where $x_k,~(k=1,2)$ are solutions of the equation
 \be  x^2-\left(\frac{m_h^2-M_1^2+M_2^2}{m_h^2}\right)x+\frac{M_2^2-i\delta}{m_h^2}=0.\label{squeq}\ee
 The $C_0$ function was given in \cite{LFVHDUgauge} consistent with that discussed on \cite{bardin}, namely
 \be  C_0=\frac{1}{m_h^2}\left[R_0(x_0,x_1)+ R_0(x_0,x_2)-R_0(x_0,x_3)\right] , \label{C0fomula1}\ee
 where
 \be R_0(x_0,x_i) \equiv Li_2(\frac{x_0}{x_0-x_i})- Li_2(\frac{x_0-1}{x_0-x_i}), \label{r0function}\ee
 $Li_2(z)$ is the di-logarithm function,  $x_{1,2}$ are solutions  of  the equation (\ref{squeq}),  and $x_{0,3}$ are given as
  \be x_0=\frac{M_2^2-M_0^2}{m_h^2},\hs x_3=\frac{-M_0^2+i\delta}{M_1^2-M_0^2}. \label{x03}\ee
 For  simplicity  of calculation we  use  approximate forms of PV functions where $p_1^2,p_2^2\rightarrow 0$, namely,
\bea  b^{(i)}_0 &=& 1-\ln\frac{M_i^2}{m_h^2}+\frac{M_0^2}{M_0^2-M_i^2}\ln\frac{M_i^2}{M_0^2},\crn
 b^{(1)}_1  &=& -\frac{1}{2}\ln\frac{M_1^2}{m_h^2}-\frac{M_0^4}{2(M_0^2-M_1^2)^2}\ln\frac{M_0^2}{M_1^2} +\frac{(M_0^2-M^2_1)(3 M_0^2-M_1^2)}{4(M_0^2-M_1^2)^2}, \crn
  b^{(2)}_1  &=& \frac{1}{2}\ln\frac{M_2^2}{m_h^2}+\frac{M_0^4}{2(M_0^2-M_2^2)^2}\ln\frac{M_0^2}{M_2^2} -\frac{(M_0^2-M^2_2)(3 M_0^2-M_2^2)}{4(M_0^2-M_2^2)^2}, \crn
b^{(12)}_0&=&\ln \frac{m_h^2-i\delta}{M_1^2-i\delta}+2 + \sum_{k=1}^2 x_k\ln\left(1-\frac{1}{x_k}\right),\crn
 C_1 &=& \frac{1}{m_h^2}   \left[b^{(1)}_0 -b_0^{(12)}+(M_2^2-M_0^2)C_0\right],\hs
  C_2 =  -\frac{1}{m_h^2}   \left[b^{(2)}_0 -b_0^{(12)}+(M_1^2-M_0^2)C_0\right]. \nn\eea
  If $M_1=M_2$, it can be seen that $b^{(1)}_1=-b^{(2)}_1$, $b^{(1)}_0=b^{(1)}_0$ and $C_1=-C_2$.

\section{\label{ffactor}Form factors for LFVHD in t' Hooft-Feynman gauge}
\begin{figure}[h]
  \includegraphics*[width=15cm]{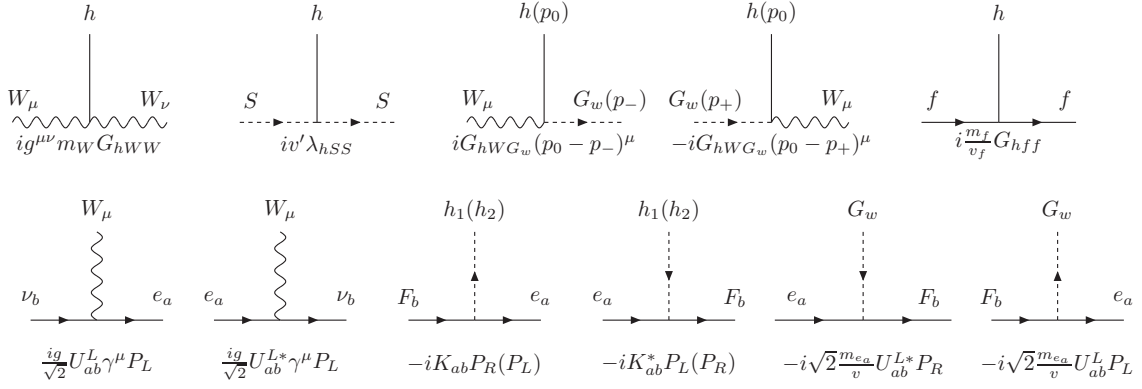}
  \caption{Feynman rules for the $h\rightarrow\mu^{\pm}\tau^{\mp}$ in the t' Hooft-Feynman gauge. Notations: i) $S=G_w,h^\pm_1,h^\pm_2$ ; ii) $K_{ab}=( y_L^TU^L)_{ab}$ and $(y_R^T)_{ab}$ for active and exotic neutrinos, respectively; iii) $f=e_a, \nu_a, N_a$, $F_a= \nu_a, N_a$. }\label{FeynRule}
\end{figure}

In this section we will list all factors for calculating LFVHD in the  model considered. The calculation is done in the t' Hooft-Feynman gauge. The Feynman rules are given in Fig. \ref{FeynRule}. These  factors  were cross-checked using FORM \cite{form}.

Contribution from  Fig. \ref{LFVdiagram} a):
  \be  i \mathcal{M}_{(1a)}=\frac{ig^2 G_{hWW}}{16\pi^2} \sum_{a=1}^3U^L_{2a}U^{L*}_{3a}\times\left( \left[\overline{u_1}P_Lv_2\right] \times E^{\nu WW}_L+ \left[\overline{u_1}P_Rv_2\right] \times E^{\nu WW}_R \right),\label{IMa}\ee
 where  $ E^{\nu WW}_L =  m_1m_W C_1 ,\hs  E^{\nu WW}_R= -  m_2 m_W C_2$ and  $C_{i}\equiv C_{i}(m_{\nu_a},m_W,m_W)$.

 Contribution from  Fig. \ref{LFVdiagram} b):
  \be  i \mathcal{M}_{(b)}=\frac{ig G_{hWG_w} }{16\pi^2}\sum_{a=1}^3U^L_{2a}U^{L*}_{3a} \times\left( \left[\overline{u_1}P_Lv_2\right]  E^{\nu W G_w}_L+ \left[\overline{u_1}P_Rv_2\right]  E^{\nu W G_w}_R\right),\label{IMb}\ee
where
\bea  E_L^{\nu W G_w}&=&-\frac{m_1m^2_{2}}{v}(2C_1-C_2), \crn
  E_R^{\nu W G_w}&=& -\frac{ m_2}{v}\left( B_0^{(12)} +m_{\nu_a}^2C_0- m_1^2C_1 +2m_2^2(-C_1+C_2)+2 m^2_{h}C_1\right) ,\; \label{E_FVS}\eea
  and $B_0^{(12)}\equiv B_0^{(12)}(m_W,m_W)$, $C_{i}\equiv C_{i}(m_{\nu_a},m_W,m_W)$.

Contribution from Fig. \ref{LFVdiagram} c):
  \be  i \mathcal{M}_{(c)}=\frac{ig G_{hWG_w}}{16\pi^2} \sum_{a=1}^3U^L_{2a}U^{L*}_{3a} \times\left(\left[\overline{u_1}P_Lv_2\right]  E^{\nu G_wW}_L+ \left[\overline{u_1}P_Rv_2\right]  E^{\nu G_wW}_R\right),\label{IMc}\ee
where
\bea   E_L^{\nu G_wW}&=& -\frac{m_1}{v}\left( B_0^{(12)} +m_{\nu_a}^2C_0 +2m_1^2(-C_1+C_2)  + m_2^2C_2-2 m^2_{h}C_2 \right),\crn
   E_R^{\nu G_wW} &=&-\frac{ m^2_{1}m_2}{v} (C_1-2C_2) , \label{E_FSV}\eea
   and $B_0^{(12)}\equiv B_0^{(12)}(m_W,m_W)$, $C_{i}\equiv C_{i}(m_{\nu_a},m_W,m_W)$.

Contribution from Fig. \ref{LFVdiagram} d):
\be  i \mathcal{M}_{(d)}=\frac{i (v' \lambda_{hSS})}{16\pi^2} \sum_{a=1}^3 V_{2a} V^{*}_{3a}\times\left(\left[\overline{u_1}P_Lv_2\right]  E^{FSS}_L+ \left[\overline{u_1}P_Rv_2\right]  E^{FSS}_R\right),\label{IMd}\ee
where  $C_{i}\equiv C_{i}(m_a,m_{S},m_{S})$, $V=\left\{U^L,K,y^T_R \right\}$, $m_a=\left\{m_{\nu_a}, m_{N_a}\right\}$ for $S=\left\{G_w, h_1, h_2\right\}$; and 
\bea    E_L^{\nu G_wG_w}&=& \frac{m_1m_2^2}{v^2}2C_2, \hs
   E_R^{\nu G_wG_w}= \frac{m_1^2m_2 }{v^2}(-2C_1), \crn
   E_L^{\nu h_1h_1}&=&-m_1C_1 , \hs
   E_R^{\nu h_1h_1}=m_2 C_2; \hs
   E_L^{Nh_2h_2}=m_2 C_2, \hs
   E_R^{Nh_2h_2}=-m_1C_1,
   \label{E_FSS}\eea

Contribution from Fig. \ref{LFVdiagram} e):
 \be  i \mathcal{M}_{(e)}=\frac{i g^2 G_{h \nu\nu}}{16\pi^2}\sum_{a=1}^3 U^L_{2a}U^{L*}_{3a} \left(\left[\overline{u_1}P_Lv_2\right]  E^{W\nu\nu}_L+ \left[\overline{u_1}P_Rv_2\right]  E^{W\nu\nu}_R\right),\label{IMe}\ee
where
\be   E_L^{W\nu\nu}=-\frac{m_1 m_{\nu_a}^2}{v'}\left(C_0-2C_1\right), \;   E_R^{W\nu\nu} =- \frac{ m_2 m_{\nu_a}^2}{v'} \left( C_0+2 C_2\right), \label{E_VFF1}\ee
and $C_{i}\equiv C_{i}(m_W,m_{\nu_a},m_{\nu_a})$.

Contribution from Fig. \ref{LFVdiagram} f):
 \be  i \mathcal{M}_{(f)}=\frac{i  G_{hFF}}{16\pi^2} \sum_{a=1}^3V_{2a}V^*_{3a}\left(\left[\overline{u_1}P_Lv_2\right]  E^{SFF}_L+ \left[\overline{u_1}P_Rv_2\right]  E^{SFF}_R\right),\label{IMf}\ee
where $C_{i}\equiv C_{i}(m_S,m_{a},m_{a})$, $V=\left\{U^L,K\right\}$ , $m_a=\left\{m_{\nu_a},m_{N_a}\right\}$ for  $S=\left\{G_w, h_1, h_2\right\}$; and
\bea   E_L^{G_w\nu\nu}&=&-\frac{m_1m_{\nu_a}^2}{v'}\left[  \frac{m^2_2 }{v^2}2(C_0+2C_2) \right],\hs
 E_R^{G_w\nu\nu}=- \frac{ m_2 m_{\nu_a}^2}{v'}\left[ \frac{m^2_1}{v^2}2(C_0-2C_1) \right], \crn
  E_L^{h_1\nu\nu} &=&-\frac{ m_1m_{\nu_a}^2}{v'}(C_0-2C_1),\hs
 E_R^{h_1\nu\nu} =- \frac{ m_2m_{\nu_a}^2}{v'}(C_0+2C_2),\crn
 E_L^{h_2NN}&=&-\frac{ m_2 m_{N_a}^2}{v'}(C_0+2C_2) ,\hs
 E_R^{h_2NN}=- \frac{ m_1 m_{N_a}^2}{v'}(C_0-2C_1),
 \label{E_SFF}\eea
Contribution from sum of Figs. \ref{LFVdiagram} g) and \ref{LFVdiagram} h):
 \be  i \mathcal{M}_{(gh)}= \frac{ig^2G_{hee}}{16\pi^2} \sum_{a=1}^3U^L_{2a}U^{L*}_{3a} \left(\left[\overline{u_1}P_Lv_2\right]  E^{\nu W}_L+ \left[\overline{u_1}P_Rv_2\right]  E^{\nu W}_R\right),\label{IMgh}\ee
where
\be   E_L^{\nu W}(m_F,m_W)= -\frac{m_1m_2^2}{(m_1^2-m_2^2)v}\left[B^{(1)}_1+B^{(2)}_1 \right], \hs E_R^{\nu W}(m_F,m_W)=\frac{m_1}{m_2}E_L^{\nu W},  \label{E_FV}\ee
 $B_1^{(i)}\equiv B_1^{(i)}(m_{\nu_a},m_W)$.

Contribution from sum of Figs. \ref{LFVdiagram} i) and \ref{LFVdiagram} k):
 \be  i \mathcal{M}_{(ik)}= \frac{i G_{hee}}{16\pi^2} \sum_{a=1}^3V_{2a}V^*_{3a}\left(\left[\overline{u_1}P_Lv_2\right]  E^{FS}_L+  \left[\overline{u_1}P_Rv_2\right]  E^{FS}_R\right),\label{IMik}\ee
where $B_1^{(i)}\equiv B_1^{(i)}(m_{a},m_S)$, $V=\left\{U^L,K,y^T_R \right\}$, $m_a=\left\{m_{\nu_a}, m_{N_a}\right\}$ for $S=\left\{G_w, h_1, h_2\right\}$; and
\bea   E_L^{\nu G_w}&=& - \frac{2 m_1^3 m_2^2 }{(m_1^2-m_2^2)v^3}\left(B^{(1)}_1+B^{(2)}_1\right) , \hs E_R^{\nu G_w}=\frac{m_2}{m_1}E_L^{\nu G_w}, \crn
E_L^{\nu h_1}&=& \frac{-m_1m_2^2 }{(m_1^2-m_2^2)v}\left(B^{(1)}_1+B^{(2)}_1\right), \hs
E_R^{\nu h_1}= \frac{m_1}{m_2}E_L^{\nu h_1},\crn
E_L^{Nh_2}&=& - \frac{ m_1^2 m_2 }{(m_1^2-m_2^2)v}\left(B^{(1)}_1+B^{(2)}_1\right) , \hs E_R^{Nh_2}=\frac{m_2}{m_1}E_L^{Nh_2},
 \label{E_FS}\eea
The divergence cancellation of the total amplitudes of the LFV decays  is proved as follows. The divergences appear only in the expressions listed in  (\ref{E_FVS}),  (\ref{E_FSV}),  (\ref{E_FV})  and  (\ref{E_FS}).  The expressions in (\ref{E_FVS}) and   (\ref{E_FSV}) will vanish after inserting them  into (\ref{IMb}) and  (\ref{IMc}), where the GIM mechanism works. The divergences in each contribution given in (\ref{E_FV}) and (\ref{E_FS}) cancel each other because $\mathrm{Div}[B^{(1)}_1]=-\mathrm{Div}[B^{(2)}_1].$   Furthermore, the limit $p_1^2,p_2^2\rightarrow 0$ results in $B^{(1)}_1=-B^{(2)}_1$; therefore all of the aforementioned contributions are  very suppressed.


\begin{thebibliography}{99}
\bibitem{higgsdicovery1}G.  Aad \emph{et al}. (ATLAS Collaboration), Phys. Lett. \textbf{B 716},  1 (2012),
arXiv:1207.7214.
 \bibitem{higgsdicovery2} V. Khachatryan \emph{et al.} (CMS Collaboration), Phys. Lett. \textbf{B 716}, 30 (2012), arXiv:1207.7235.

 \bibitem{exLFVh}V. Khachatryan \emph{et al.} (CMS Collaboration), Phys. Lett. \textbf{B 749}, 337  (2015); G. Aad \emph{et al.} (ATLAS Collaboration), JHEP \textbf{1511} (2015) 211, arXiv: hep-ex/1508.03372.

 \bibitem{LFVhtauemu} CMS Collaboration, (2015), CMS-PAS-HIG-14-040, http://inspirehep.net/record/1388810/

 \bibitem{seesaw} A. Pilaftsis, Phys.  Lett. \textbf{B 285}, 68 (1992); J. G. K\"{o}rner, A. Pilaftsis, K. Schilcher, Phys. Rev. D {\bf 47} (1993) 1080;  E. Arganda, A. M. Curiel, M. J. Herrero, D. Temes, Phys. Rev. D \textbf{71}, 035011  (2005), arxiv:  hep-ph/0407302.

\bibitem{iseesaw}E. Arganda, M. J. Herrero, X. Marcano and C. Weiland, Phys. Rev. D \textbf{91}, 015001 (2015).

\bibitem{SUSY} A. Brignole,  A.  Rossi, Phys. Lett. \textbf{B 66}, 217  (2003), arXiv:hep-ph/0304081; A. Brignole, A. Rossi, Nucl. Phys.  \textbf{B  701}, 3 (2004), arXiv:hep-ph/0404211; M. Arana-Catania, E. Arganda, M. J. Herrero, JHEP \textbf{1309}, 160 (2013);  JHEP \textbf{1510},  192 (2015);
E. Arganda, M. J. Herrero, X. Marcano, C. Weiland,  Phys. Rev. \textbf{D 93}, 055010 (2016) , arXiv: hep-ph/1508.04623;  P. T. Giang, L. T. Hue, D. T. Huong and H. N. Long, Nucl. Phys. \textbf{B 864} (2012) 85, [arXiv:1204.2902(hep-ph)]; L. T. Hue, D. T. Huong, H. N. Long, H. T. Hung, N. H. Thao, Prog. Theor. Exp. Phys.  \textbf{113B05} (2015), arXiv:1404.5038 [hep-ph];  D. T. Binh, L. T. Hue, D. T. Huong, H. N. Long, Eur. Phys. J. \textbf{C 74} (2014) 2851, [arXiv:1308.3085(hep-ph)];  E. Arganda, M. J. Herrero, R. Morales and A. Szynkman, JHEP \textbf{1603}, 055  (2016), arxiv: hep-ph/1510.04685; J. L. Diaz-Cruz, JHEP \textbf{0305}, 036  (2003) arXiv: hep-ph/0207030].
\bibitem{THDL1}A. Crivellin, G. D'Ambrosio, J. Heeck, Phys. Rev. Lett. \textbf{114} (2015) 151801; N. Bizot, S. Davidson, M. Frigerio, J. L. Kneur, JHEP \textbf{1603} (2016) 073.

 \bibitem{THDL2}N. Bizot, S. Davidson, M. Frigerio, J. -L. Kneur, JHEP \textbf{1603} (2016) 073; F. J. Botella, G. C. Branco, M. Nebot, M. N. Rebelo, Eur. Phys. J. \textbf{C 76} (2016), 161; S. Kanemura, T. Ota, T. Shindou and K. Tsumura, Phys. Rev. D  \textbf{73}, 016006  (2006), hep-ph/0505191; M. Arroyo, J. L. Diaz-Cruz, E. Diaz and J. A. Orduz-Ducuara, arXiv:  hep-ph/1306.2343;    D. Das and A. Kundu, Phys. Rev. D \textbf{92}, 015009  (2015), arXiv: hep-ph/1504.01125; X. F Han, L. Wang, J.  M. Yang, Phys. Lett. \textbf{B 757} (2016) 537.

 \bibitem{LFVHDUgauge} L. T. Hue, H. N. Long, T. T. Thuc and T. Phong Nguyen, Nucl. Phys. \textbf{B 907} (2016) 37,  arXiv: hep-ph/1512.03266.

\bibitem{NonSUSY} S. Baek, Z.-F. Kang, JHEP \textbf{1603} (2016) 106; S. Baek, K. Nishiwaki , Phys. Rev. D \textbf{93} (2016), 015002.

\bibitem{leptoquark}K. Cheung, W. Y. Keung, P. Y. Tseng, Phys. Rev. D \textbf{93} (2016), 015010.

 \bibitem{LFVgeneral} W. Altmannshofer, S. Gori, A. L. Kagan, L. Silvestrini, J. Zupan, Phys. Rev. D \textbf{93} (2016), 031301; X. G. He, J. Tandean, Y. J. Zheng, JHEP \textbf{1509} (2015) 093 ; I. Dor\v{s}ner, S. Fajfer, A. Greljo, J. F. Kamenik, N. Ko\v{s}nik, Ivan Ni\v{s}and\v{z}ic, JHEP \textbf{1506} (2015) 108;  R. Harnik, J. Kopp and J.  Zupan, JHEP \textbf{1303}, 026 (2013), arXiv: hep-ph/1209.1397;  A. Celis, V. Cirigliano and E. Passemar, Phys. Rev. D \textbf{89}, 013008  (2014), arXiv: hep-ph/1309.3564;  A. Dery, A. Efrati, Y. Nir, Y. Soreq and V. Susi, Phys. Rev. D \textbf{90}, 115022  (2014),  arXiv: hep-ph/1408.1371; J. Heeck, M. Holthausen, W. Rodejohann and Y. Shimizu, Nucl. Phys. \textbf{B  896}, 281  (2015),  arXiv: hep-ph/1412.3671;  A. Crivellin, G. DAmbrosio and J. Heeck, Phys. Rev. D 91, 075006 (2015), arXiv: hep-ph/1503.03477;  J. L. Diaz-Cruz and J. J. Toscano, Phys. Rev. D 62, 116005  (2000), arXiv: hep-ph/9910233;  A. Pilaftsis, Z. Phys. \textbf{C 55}, 275 (1992) ; L. D . Lima, C. S. Machado, R. D. Matheus, L. A. F.  D. Prado, JHEP \textbf{1511}, 074  (2015), arXiv: hep-ph/1501.06923;  I. d. M. Varzielas, O. Fischer, V. Maurer, JHEP \textbf{1508}, 080 (2015);  C. F. Chang, C. H. V. Chang, C. S. Nugroho, T. C. Yuan, "\emph{Lepton Flavor Violating Decays of Neutral Higgses in Extended Mirror Fermion Model }", e-Print: arXiv:1602.00680; C. H. Chen, T. Nomura, "\emph{Bound on LFV Higgs decays in a vectorlike lepton model and search for doubly charged lepton at the LHC }", arXiv:1602.07519 [hep-ph]; K. Huitu, V. Keus, N. Koivunen, O. Lebedev, JHEP \textbf{1605} (2016) 026, arXiv:1603.06614.

\bibitem{lfvNH}M. Sher, K. Thrasher,  Phys. Rev. D \textbf{93}, 055021 (2016).

\bibitem{LFVcol}S. Kanemura, K. Matsuda, T. Ota, T. Shindou, E. Takasugi and K.Tsumura, Phys. Lett. \textbf{B 599}, 83 (2004), hep-ph/0406316; G. Blankenburg, J. Ellis and G. Isidori, Phys. Lett. \textbf{B 712}, 386    (2012), arXiv:hep-ph/1202.5704;   S. Davidson and P. Verdier, Phys. Rev. D \textbf{86}, 111701 (2012), arXiv:  hep-ph/1211.1248;  S. Bressler, A. Dery and A. Efrati, Phys. Rev. D \textbf{90}, 015025 (2014),  arXiv:  hep-ph/1405.4545;  D. Aristizabal Sierra and A. Vicente, Phys. Rev. D \textbf{90}, 115004  (2014), arXiv: hep-ph/1409.7690; C. X. Yue, C. Pang and Y. C. Guo, J. Phys. \textbf{G 42}, 075003  (2015), arXiv:  hep-ph/1505.02209; S. Banerjee, B. Bhattacherjee, M. Mitra, M. Spannowsky, \emph{"The Lepton Flavour Violating Higgs Decays at the HL-LHC and the ILC "},  arXiv:1603.05952 [hep-ph]; I. Chakraborty, A. Datta, A. Kundu, \emph{"Lepton flavor violating Higgs boson decay $h\rightarrow\mu\tau$ at the ILC "}, arXiv:1603.06681 [hep-ph].

\bibitem{NuOcs} Y. Fukuda, et al., Phys. Rev. Lett. \textbf{81}, 1562 (1998); S. Fukuda et al., Phys. Rev. Lett. \textbf{85} (2000) 3999.

\bibitem{Kenji}K. Nishiwaki, H. Okada, and Y. Orikasa, Phys. Rev. D \textbf{92}, 093013 (2015).

\bibitem{Kane}S. Kanemura, K. Nishiwaki, H. Okada, Y. Orikasa, S. C. Park, R. Watanabe, "LHC 750 GeV Diphoton excess in a radiative seesaw model", arXiv:1512.09048 [hep-ph].

\bibitem{actnuUpdate} M. C. Gonzalez-Garcia, M. Maltoni, T. Schwetz,  JHEP \textbf{1411} (2014) 052 .

\bibitem{spinor} H. K. Dreiner, H. E. Haber, S. P. Martin,  Phys. Rept. \textbf{494},  1 (2010).

 \bibitem{bardin} D. Y. Bardin, G. Passarino, "{\it The Standard Model in the making: Precision study of the electroweak interactions}", Clarendon Press-Oxford, 1999.

 \bibitem{PDG2014}  K. A. Olive et al. (Particle Data Group), Chin. Phys. C \textbf{38},  090001 (2014).

 \bibitem{yLbound} J.  Herrero-Garcia, M. Nebot, N. Rius, and A. Sntamaria, Nulc. Phys. B \textbf{885}, 542 (2014).

 \bibitem{form} J. A. M. Vermaseren, "\emph{New features of FORM} ", arxiv: math-ph/0010025.

\bibitem{cHseach} ATLAS Collaboration (G. Aad \emph{et al}.), JHEP \textbf{1206}, 039  (2012); ATLAS Collaboration (G. Aad \emph{et al}.), JHEP \textbf{1603} (2016) 127.

\bibitem{Hooft} G. 't Hooft and M. Veltman, Nucl. Phys. \textbf{B 153}, 365  (1979).

\bibitem{m3loop}T. Nomura, H. Okada, Phys. Lett. \textbf{B 755} (2016) 306; T. Nomura, H. Okada, \emph{"Four-loop Radiative Seesaw Model with 750 GeV Diphoton Resonance"}, arXiv:1601.04516 [hep-ph].
\bibitem{Nueecolider} S. Antusch, E. Cazzto, O. Fischer, JHEP \textbf{1604}, 189 (2016); S. Antusch, O. Fischer, JHEP \textbf{1505} (2015) 053;  JHEP \textbf{1410}, 094 (2014).
\bibitem{Gu1}Y. Mambrini, S. Profumo, and F. S. Queiroz, \emph{"Dark Matter and Global Symmetries"},   arXiv:1508.06635v1 [hep-ph].
\end{thebibliography}
\end{document}